\documentclass[12pt]{iopart}
\usepackage{epsfig}
\usepackage{graphicx}
\usepackage{epstopdf}
\usepackage[section]{placeins}
\usepackage{color}
\usepackage{cite}

\begin{document}
\title{Magnetomechanical response of bilayered magnetic elastomers} 

\author{Elshad Allahyarov}
\address{
 Institut f\"ur Theoretische Physik II, Heinrich-Heine-Universit\"at  D\"usseldorf
  Universit\"atstrasse 1, 40225 D\"usseldorf, Germany \\
 Theoretical Department, Joint Institute for High Temperatures, Russian Academy of Sciences (IVTAN),
13/19 Izhorskaya street, Moscow 125412, Russia \\
 Department of Macromolecular Science and Engineering, Case Western Reserve University, 
  Cleveland, Ohio 44106-7202, United States
}

\author{Andreas M.\ Menzel}
\address{ 
Institut f\"ur Theoretische Physik II, Heinrich-Heine-Universit\"at  D\"usseldorf
  Universit\"atstrasse 1, 40225 D\"usseldorf, Germany  \\
}

\author{Lei Zhu}
\address{Department of Macromolecular Science and Engineering, Case Western Reserve University, 
  Cleveland, Ohio 44106-7202, United States}

\author{Hartmut L\"owen}
\address{ 
Institut f\"ur Theoretische Physik II, Heinrich-Heine-Universit\"at  D\"usseldorf
  Universit\"atstrasse 1, 40225 D\"usseldorf, Germany  \\
}

\ead{elshad.allakhyarov@case.edu}

\begin{abstract}

Magnetic elastomers are appealing materials from an application point of view: they combine the mechanical softness and 
deformability of polymeric substances with the addressability by external magnetic fields. 
In this way, mechanical deformations can be reversibly induced and elastic moduli can be reversibly adjusted from outside. 
So far, mainly the behavior of single-component magnetic elastomers and ferrogels has been studied. 
Here, we go one step further and analyze the magnetoelastic response of a bilayered material 
composed of two different magnetic elastomers. 
It turns out that, under appropriate conditions, the bilayered magnetic elastomer can show a 
strongly amplified deformational response in comparison to a single-component material. 
Furthermore, a qualitatively opposite response can be obtained,
 i.e.\ a contraction along the magnetic field direction (as opposed to an elongation in 
the single-component case). 
We hope that our results will further stimulate experimental and theoretical investigations 
directly on bilayered magnetic elastomers, or, in a further hierarchical step, on bilayered units embedded in yet another polymeric matrix. 

 \end{abstract}
\maketitle

\section{Introduction}

The terms ``magnetic hybrid materials'' or ``magnetic composite materials'' are typically associated with 
classical magnetic elastomers or ferrogels \cite{filipcsei2007magnetic}. These substances consist of a 
more or less chemically crosslinked and possibly swollen polymeric matrix into which paramagnetic, 
superparamagnetic, or ferromagnetic colloidal particles are embedded. In this way, the advantageous features 
of two different classes of materials are combined into one: on the one hand, one obtains free-standing soft 
elastic solids of typical polymeric properties \cite{doi1988theory,strobl1997physics}; on the other hand, the
 materials can be addressed by external magnetic fields and in this way their properties can be tuned 
reversibly from outside as for conventional ferrofluids and magnetorheological fluids 
\cite{rosensweig1985ferrohydrodynamics,odenbach2003ferrofluids,odenbach2003magnetoviscous,
huke2004magnetic,odenbach2004recent,fischer2005brownian,ilg2005structure, embs2006measuring,ilg2006structure,gollwitzer2007surface,vicente2011magnetorheological}.

A lot of work has been spent on investigating how such ferrogels mechanically respond to external magnetic fields. 
In particular, these analyses focused on the nature of the induced shape changes \cite{zrinyi1996deformation,zrinyi1997direct,filipcsei2007magnetic, guan2008magnetostrictive,filipcsei2010magnetodeformation, ivaneyko2011magneto,wood2011modeling,camp2011effects,stolbov2011modelling, zubarev2012theory,ivaneyko2012effects,weeber2012deformation,gong2012full, zubarev2013effect,zubarev2013magnetodeformation,ivaneyko2014mechanical}. 
It turned out that the spatial distribution of the magnetic particles within a sample can qualitatively
 influence its response to the external field. This is because the magnetic interaction between the magnetic
 particles depends on their spatial arrangement. For example, when the particles were arranged on regular lattice structures, 
the system showed either an elongation along the external field or a contraction, depending on the particular lattice 
\cite{ivaneyko2011magneto,ivaneyko2012effects,ivaneyko2014mechanical}. Likewise, anisotropic particle distributions 
and the presence of chain-like aggregates that can for example result from crosslinking the polymer matrix in the 
presence of a strong external magnetic field \cite{collin2003frozen,varga2003smart,gunther2012xray,borbath2012xmuct,gundermann2013comparison} 
can change the mechanical properties in and the deformational response to an external magnetic 
field \cite{bohlius2004macroscopic,filipcsei2007magnetic,filipcsei2010magnetodeformation, wood2011modeling, han2013field,zubarev2013effect,zubarev2013magnetodeformation}.
 Even the presence of randomly distributed dimer-like arrangements instead of single isolated
 magnetic particles was shown to be able to switch the distortion of a ferrogel from contractile 
along the field direction to extensile \cite{stolbov2011modelling,gong2012full}.

The coupling of mechanical and deformational behavior to external magnetic fields, often referred to as ``magnetomechanical coupling'', 
opens the way to various different types of application. Soft actuators \cite{zimmermann2006modelling} or magnetic sensors 
\cite{szabo1998shape,ramanujan2006mechanical} can be constructed that react mechanically to external magnetic fields or 
field gradients. Vibration absorbers \cite{deng2006development} and damping devices \cite{sun2008study} can be manufactured, 
the properties of which can be reversibly tuned from outside by applying an external magnetic field. In the search for an 
increased magnitude of magnetomechanical coupling, a new class of ferrogels was synthesized \cite{frickel2011magneto,messing2011cobalt}. 
Via surface functionalization of the magnetic particles, the polymer chains could be directly chemically attached to the particle 
surfaces. In this way, the magnetic particles became part of the embedding crosslinked polymer network. For such materials, a 
restoring mechanical torque acts on the particles when they rotate out of their initial orientations acquired during crosslinking 
\cite{weeber2012deformation}, which constitutes a form of orientational memory \cite{annunziata2013hardening}.

Here we analyze a further type of magnetomechanical coupling that arises when layered materials of magnetic elastomers are 
considered. In our case, we investigate the deformational response of a bilayered magnetic elastomer to an external magnetic 
field combining phenomenological magnetostatics with elasticity theory. This 
type of deformation is related to the magnetic pressure on the 
material boundaries, similar to the Maxwell pressure in dielectrics. 
The boundary-related magnetic pressure acts not only on the 
outer surfaces of the sample, but also on the inner interfaces in the considered composite materials 
containing two (or many) layers of magnetic elastomers of different magnetic susceptibilities \cite{mullins-1956}. 
In conventional ferromagnetic materials this setup was for example suggested for sensor applications when two ferromagnetic prisms are 
separated by a piezolayer \cite{liverts-2011}.

In this paper we  theoretically analyze  basic principles of the deformation 
 in composite magnetic elastomers generated by the magnetic pressure of external fields. 
We connect the ultimate deformation of the composite material to the effective magnetic pole distribution on the 
 material boundaries and at the bilayer interface. The material properties of the composite elastomer, such as its 
susceptibility and demagnetization coefficients, define a crucial parameter, called 
a geometrical function $A$, which plays a major role in the material reaction to 
the applied field. It turns out that, under appropriate conditions, the bilayered magnetic elastomer can show a 
strongly amplified deformational response in comparison to a single-component material. 
Furthermore, a qualitatively opposite response can be obtained,
 i.e.\ a contraction along the magnetic field direction (as opposed to an elongation in 
the single-component case). 

The rest of the paper is organized as follows. 
In the next section the magnetic pressure on the particle boundaries and the geometry function $A$ are 
 explained. The dependence of the function $A$ on the demagnetization coefficient $\alpha$ is discussed in section 3. 
In section 4 we investigate  the magnetization of the different layers in the bilayer. 
 The magnetic pole distribution  and the expression for 
the elastic strain are discussed in section 5.  Finally we conclude in section 6.

\clearpage
\newpage

\section{Magnetic energy density and magnetic driving pressure}
When a homogeneous material with a magnetic permittivity 
$\mu$ is placed into a uniform external magnetic field $\vec H_0$, 
the driving pressure  on the material boundary is associated with the
difference between the  
relative energy densities  
\begin{equation}
 \Delta u = \frac{\mu H^2}{2} - \frac{\mu_0 H_0^2}{2}. 
\label{single-layer-energy}
\end{equation}
Here $\mu_0$ is the susceptibility of vacuum, $\mu_0 =  1.26 \times10^{-6}$ $\frac{m kg}{s^2 A^2}$, 
and $H$ denotes the internal magnetic field. 
Using the relation between the magnetization $M$ and the internal field $H$ in an isotropic material, 
\begin{equation}
M = \chi H, 
\end{equation} 
where $\chi$ is the susceptibility of the material, and 
\begin{equation}
H=\frac{H_0}{1 + \alpha \chi}
\label{alpha}
\end{equation}
is the internal field in the material. Here, $\alpha$ is the demagnetization coefficient of the sample along the 
field direction $\vec z$. Assuming a rectangular   prism geometry 
for the material,  the full energy difference can be rewritten as 
\begin{equation}
\Delta U =  V \Delta u  =  \frac{\mu_0 M H_0}{2} A(\alpha,\chi) \, V.
\label{single-layer-energy-1}
\end{equation}
Here $V=L^2 L_z$ is the volume of the material, 
with $L_z$ the edge length of the sample along the field direction $\vec z$, and $L$ the edge lengths in 
the remaining lateral directions set equal for simplicity.
 The demagnetization coefficient $\alpha$ depends on the dimensional lengths of the sample. 
In general, the factor $\alpha^{(l)}$ has $l=x, y, z$ components, 
which obey $\sum_l \alpha^{(l)} =1$. 
Whereas for simple geometrical shapes  the coefficients 
$\alpha^{(l)}$ are well known, for example, for a sphere and a cube 
 $\alpha^{(x)} = \alpha^{(y)} =\alpha^{(z)} =1/3$, and for a slab with infinite lateral $(xy)$ dimensions
$\alpha^{(x)} = \alpha^{(y)} =0$, and $\alpha^{(z)} =1$, 
for the rectangular prisms considered in this work 
 the coefficients $\alpha^{(l)}$ are not known {\it apriori}. Their values, however, 
can be calculated using analytical expressions given in Ref.~\cite{aharoni-1998,aharoni-2000}, or 
taken from the tabulated results available in Ref.~\cite{chen-2005}.   
In the following we will adopt the notion $\alpha = \alpha^{(z)}$.

The geometry factor $A(\alpha,\chi)$ in Eq.(\ref{single-layer-energy-1}) turns out to be
\begin{equation}
 A(\alpha,\chi) =  \frac{1-2 \alpha - \alpha^2 \chi}{1+ \alpha \chi}
\label{single-layer-energy-2}
\end{equation}
and obeys $\vert A(\alpha,\chi)\vert \le 1$. This function defines 
the type of the deformation: a positive $A(\alpha,\chi)$ means a stretching, and a negative  $A(\alpha,\chi)$ 
 means a compression of the material along the applied field. A full description of this function is presented in  section~\ref{section-geometry}.

Under the driving pressure the material is deformed because of the propagation of the
material boundary from the area with  high susceptibility into  the surrounding area with low susceptibility.
The deformational changes of the material, namely the  changes in its thickness
 $L_z$ and area $S=L^2$, lead to the magnetic energy difference in Eq.(\ref{single-layer-energy-1})
\begin{equation}
\Delta U = \frac{\partial U} {\partial L_z} \Delta L_z + \frac{\partial U} {\partial S} \Delta S.
\label{single-force-4}
\end{equation}

Assuming that the density of the  material does not change during this type of deformation, 
which is referred to as a constant  volume condition, written as 
\begin{equation}
S L_z = (S+ \Delta S) (L_z-\Delta L_z),
\label{constvol}
\end{equation} 
we get the following relation between $\Delta S$ and $\Delta L_z$ when $\Delta L_z/L_z \ll$1
\begin{equation}
\Delta S = S \frac{\Delta L_z}{ L_z}
\label{single-force-8}
\end{equation}  

From  Eq.(\ref{single-layer-energy-1}), assuming that neither 
$ M$ nor $ H$ strongly depend on the changes in the material geometry, 
which is valid only for small shape deformations, for the partial derivatives of the stored energy 
we find 
\begin{eqnarray}
\frac{\partial U} {\partial L_z}  =  \frac{1}{2}  \mu_0 H_0 M S  A(\alpha,\chi) \,\, ,  \nonumber \\
\frac{\partial U} {\partial S}  =  \frac{1}{2}  \mu_0 H_0 M L_z  A(\alpha,\chi).
 \label{single-force-9}
\end{eqnarray}

Inserting Eq.(\ref{single-force-9}) into   Eq.(\ref{single-force-4}) and using Eq.(\ref{single-force-8}) we obtain
\begin{equation}
\Delta U = \mu_0 H_0 M S \Delta  L_z  \,  A(\alpha,\chi).
\label{single-force-11}
\end{equation}
Taking into account the force-energy relation 
 $F =  \Delta U / \Delta L_z$, we obtain the final expression for the magnetic pressure,
\begin{equation}
p = \vert \vec F\vert / S =  \mu_0 \vert \vec H_0 \vert  A(\alpha,\chi) (\vec M \cdot \vec n) .
\label{single-force-12}
\end{equation}
Here $\vec n$ is a unit vector normal to the  material surface 
pointing outward the sample surface. It should be noted that for specific cases 
when the constant volume condition  Eq.(\ref{constvol}) does not apply, for example, in magnetic liquids 
which can leave the field area when squeezed by magnetic forces, the second term containing $\Delta S$ in 
 Eq.(\ref{single-force-4}) can be zero. For such cases the magnetic pressure will be half 
the pressure defined by  Eq.(\ref{single-force-12}).

The term  $\vec M \cdot \vec n$ in  Eq.(\ref{single-force-12}) defines the effective magnetic pole density \cite{jackson1998classical} 
\begin{equation}
\sigma_M = \vec M \cdot \vec n
\end{equation}
at the boundaries of the sample. The sign of these magnetic poles is positive on the 
upper boundary, and negative on the bottom boundary of the sample if the field direction is 
from the bottom to the top.

\clearpage
\newpage
\section{Geometry-dependent deformation under an applied field}
\label{section-geometry}

It is evident 
from  Eq.(\ref{single-force-12}) that  the 
 magnetic driving force $\vec F$ acts on the upper and bottom surfaces in opposite directions trying to 
stretch the prism if $A(\alpha,\chi)$ 
is positive. In the opposite case, when $A(\alpha,\chi)$ is negative, the driving force $\vec F$
pushes the upper and bottom boundaries towards each other. 

\begin{figure}  [!h]
\begin{center}
\includegraphics*[width=0.45\textwidth,height=0.45\textwidth]{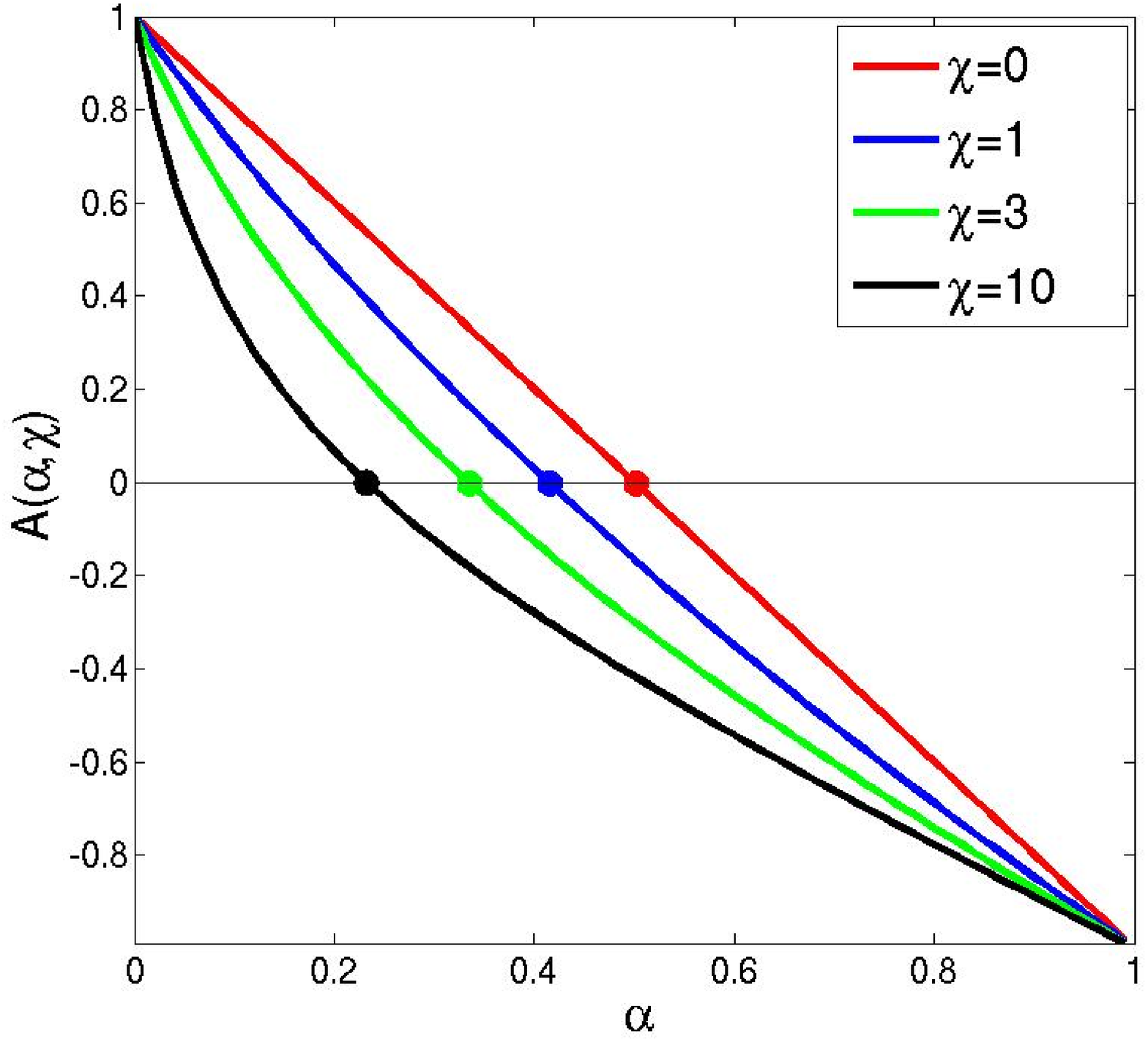}
\end{center}
\caption{(Color in online)  
 Geometrical factor $A(\alpha,\chi)$ for four different values of the magnetic susceptibility $\chi$. 
Note that the zeros of $A$ correspond to the particular dimensions of the prism at which no deformation
of the material is observed.
 \label{fig-A-factor}
}
\end{figure}

The limiting boundaries of $A(\alpha,\chi)$ are dictated by the dependence of the
coefficient $\alpha$ on the geometry of the material. For an infinite slab  with $L_z/L \to 0$ and
$\alpha \approx 1$ 
 the function $A$ reaches its bottom limit $A(\alpha,\chi) \approx -1$  from  Eq.(\ref{single-layer-energy-2}), 
which recovers the classical relation $u=-\mu_0 M H_0 /2$ for the magnetic energy density \cite{sheikh-2003}.
Putting $\alpha=0$ (the case of an elongated cylinder along the $z$-axis) into   Eq.(\ref{single-layer-energy-2})
we get the upper limit for $A(\alpha,\chi) = 1$.  The function  $A(\alpha,\chi)$ changes its sign 
 at $\alpha=\sqrt{1+\chi} -1$, as shown in Figure~\ref{fig-A-factor}. 
The zeros of $A$ correspond to the particular dimensions of the prism at which no deformation
of the material is observed.

\begin{figure}  [!h]
\begin{center}
\includegraphics*[width=0.45\textwidth,height=0.45\textwidth]{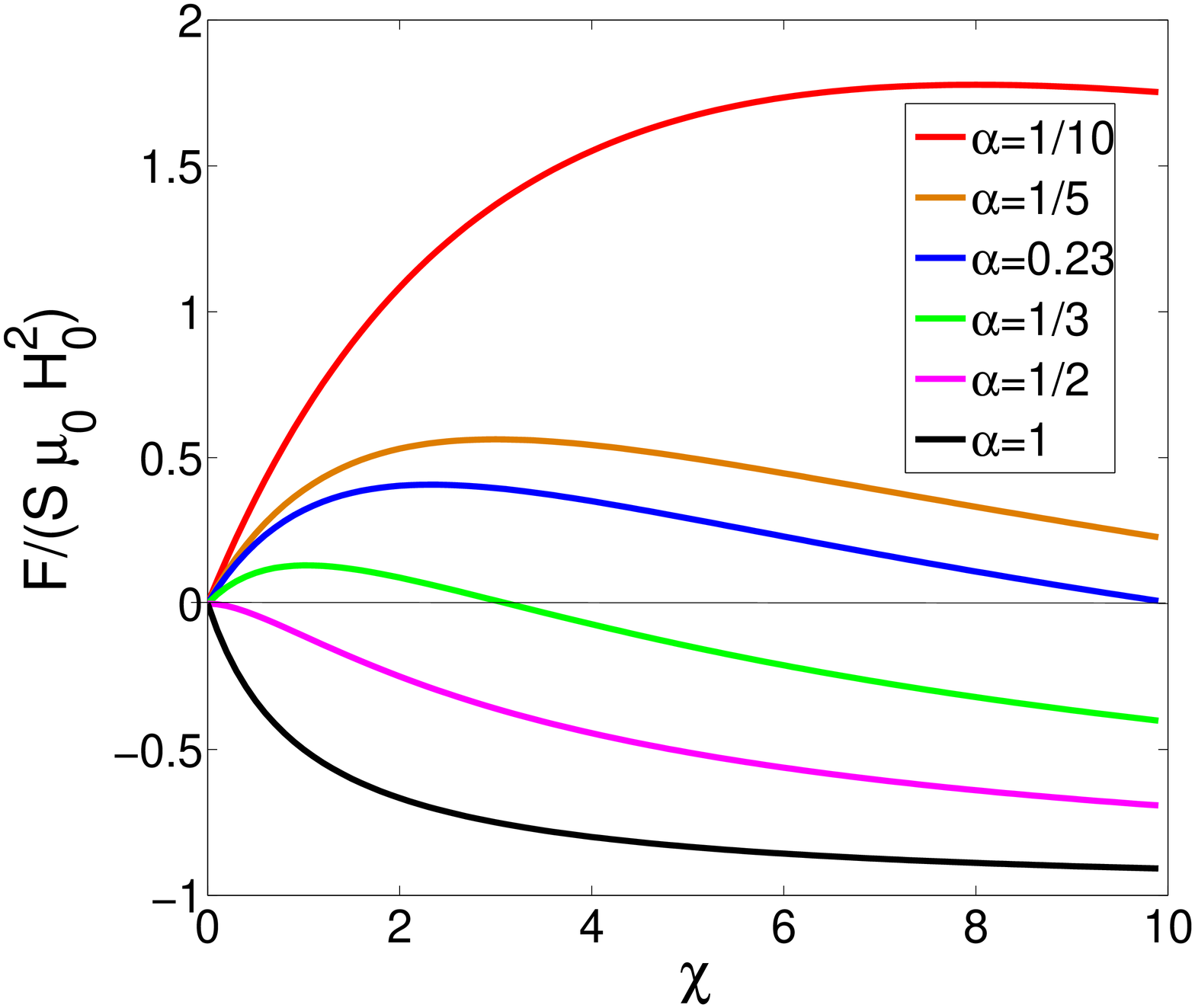}
\end{center}
\caption{(Color in online)  
Normalized pressure $F/(S \mu_0 H_0^2)$ as a function of the magnetic susceptibility  $\chi$ for 
 six different values of $\alpha$. Note that the force becomes completely positive  at 
$\alpha < 0.23$, and completely negative at  $\alpha >0.5$. 
$F>0$ corresponds to an elongation along the magnetic field direction, whereas $F<0$ implies a contraction. 
 \label{fig-FC-factor}
}
\end{figure}

For the case  $0< \chi<10$ considered in this paper, 
the function $A(\alpha,\chi)$  is always positive for $\alpha < 0.23$, which corresponds roughly to the size ratio $L_z/L > 3.3 $.
An opposite scenario, a shrinking of the prism for all $0< \chi<10$ is predicted for $\alpha> 0.5$ 
which roughly corresponds to the size ratio  $L_z/L < 1.4  $.  This is demonstrated in Figure~\ref{fig-FC-factor} where the magnetic pressure $F/S$
is plotted against the susceptibility $\chi$.

\clearpage
\newpage

\section{Magnetization of a magnetic bilayer under external field}
\begin{figure}  [!ht]
\begin{center}
\includegraphics*[width=0.6\textwidth,height=0.4\textwidth]{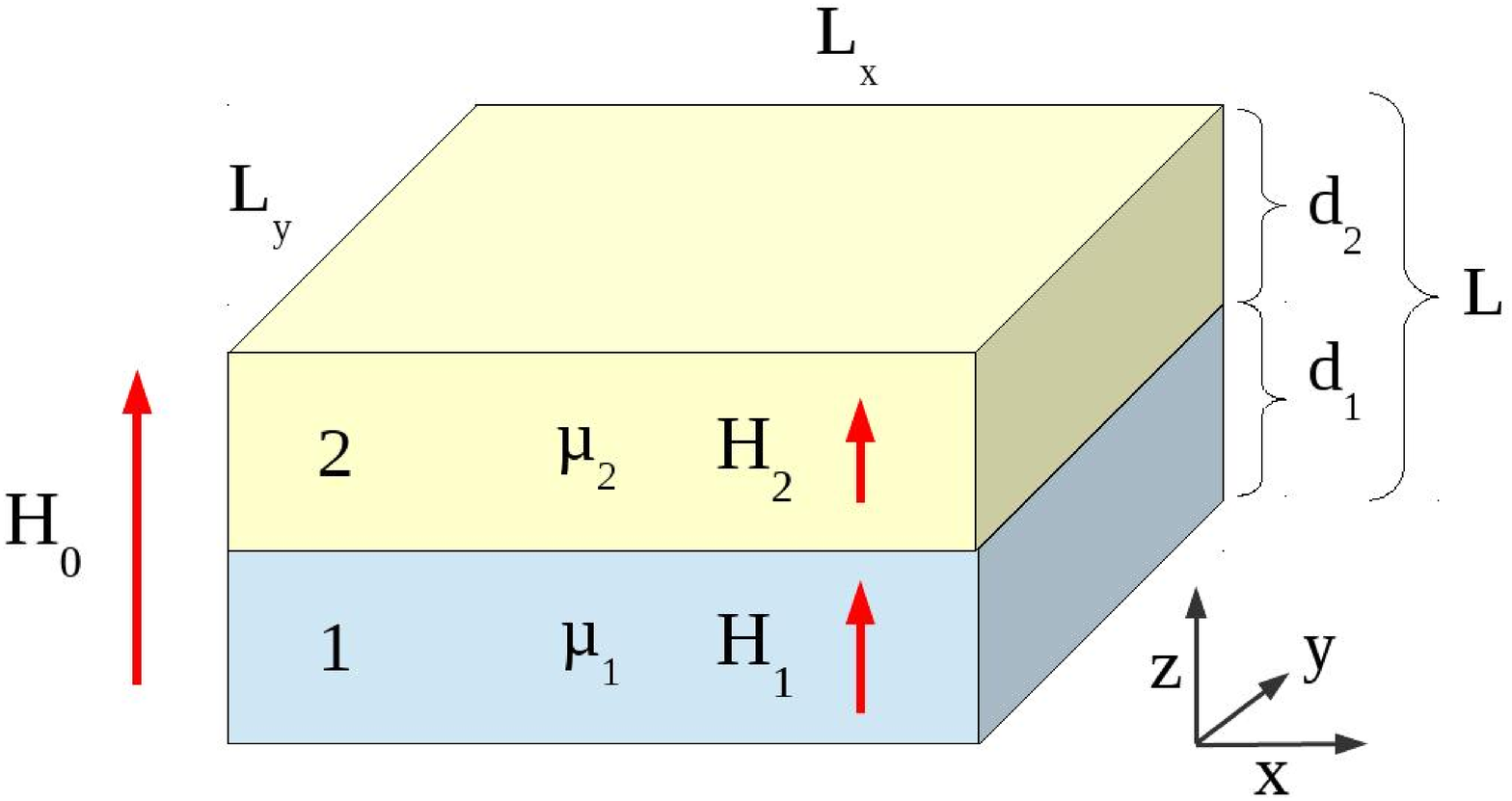}
\end{center}
\caption{(Color in online)  Schematic illustration of the composite 
bilayered magnetic elastomer in the form of a prism with 2-2
connectivity under an external magnetic field $H_0$. 
 \label{fig-1}
}
\end{figure}

We now consider a composite bilayered magnetic elastomer  of  
a rectangular shape with a 2-2 connectivity \cite{bowen-2003} as shown in Figure~\ref{fig-1}.  
The rectangular prism has  dimensions $L_x, L_y, L_z$, and for simplicity we assume that 
 its lateral dimensions are the same,  $L_x=L_y=L$. 
The bottom and upper parts of the prism, denoted as layers  {\it i=1} and {\it i=2}, 
are made from different materials with magnetic susceptibilities $\mu_1$ and $\mu_2$,
 and elastic moduli $Y_1$ and $Y_2$, and have thicknesses $d_1$ and $d_2=L_z-d_1$ correspondingly. 
There is no gap between the layers of the prism, $d=0$, hence the
stacking density of the composite is $\rho = L_z/(d+L_z) = 1$.

When an external magnetic field $\vec B_0 = \mu_0 \vec H_0$ is applied along the $z$-axis, 
$\vec H_0 \parallel \vec z$, the field $\vec B_i$ in the layer $i$ is determined as 
\begin{equation}
 \vec B_i = \mu_0 ( \vec H_i + \chi_i  \vec H_i)
\label{eq-10a}
\end{equation}
where the magnetic field $ \vec H_i$  is defined as 
\begin{equation}
\vec H_i = \vec H_0  +  \vec H_{d}^{(i)}  + \vec R_i(\vec H_j).
\label{eq-1}
\end{equation} 
Here $ \vec H_{d}^{(i)}$ is a demagnetization field in the prism $i$ \cite{smith-2010}.  
 This field originates from the existence of 
 magnetic poles at the boundaries of the prism $i$ perpendicular to the field, in 
analogy to the polarization charges at the 
dielectric boundaries under an external electric field. In the linear response theory the 
field  $ \vec H_{d}^{(i)}$ reads 
\begin{equation}
\vec H_{d}^{(i)} = - \alpha_i \vec M_i,
\label{eq-2}
\end{equation} 
where $\alpha_i$ is the demagnetization factor of the prism $i$ along the $z$-axis.

\begin{figure}  [!ht]
\begin{center}
\includegraphics*[width=0.8\textwidth,height=0.6\textwidth]{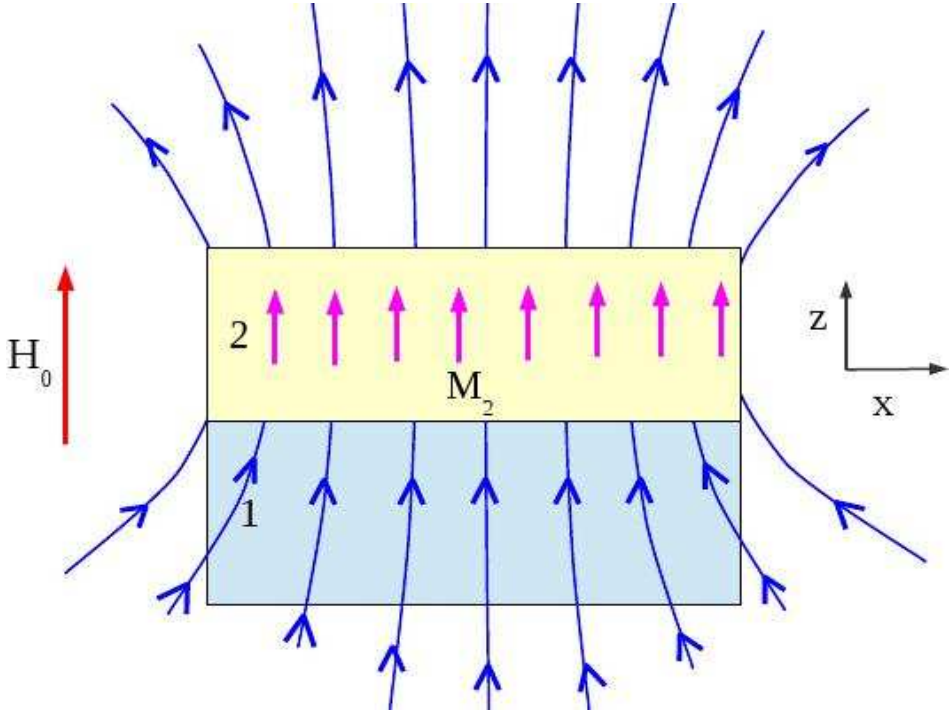}
\end{center}
\caption{(Color in online)  Schematic presentation of the cross term $\vec R_1(\vec H_2)$, 
which corresponds to the field lines generated by $\vec M_2$ of layer {\it 2} in 
the volume of layer {\it 1}. 
 \label{fig-2}
}
\end{figure}

The last term on the right hand side of Eq.(\ref{eq-1}), $\vec R_i(\vec H_j)$, 
represents the average value of the magnetic field $\vec H_j$ generated by the magnetized layer $j$ in the 
volume of layer $i$, where $j\neq i$. 
The full distribution of this cross field can be  calculated using numerical methods, 
see Ref.~\cite{christensen-2011}. The field $\vec R_1(\vec H_2)$, 
schematically drawn in Figure~\ref{fig-2}, is inhomogeneous along the $z$-axis: it has a maximum value at the 
  top of the layer {\it 1} and becomes weaker towards the bottom edge of the layer {\it 1}.
There are different approaches about accepting the best approximation 
for $R_i$, see Ref.~\cite{aharoni-2002}. The so-called 'ballistic'
approach defines $\vec R_i$ as the averaged $\vec R_i(\vec H_2)$ in the $xy$ mid-plane of layer {\it 1}. 
Or, the 'local' approach defines $\vec R_i(\vec H_2,z)$ along the central line $z$ with $x=y=0$. 
Within the 'side' approach $\vec R_i$ is measured as an averaged field over the surfaces of the layer $i$
 perpendicular to $z$.
 In our generalized approach 
we assume 
 that the average field $\vec R_i$ is homogeneous across the layer $i$  and is a fraction of the magnetization of 
layer {\it j}, 
\begin{equation}
 \vec R_i  = \gamma_i \vec M_j.
\label{eq-2a}
\end{equation}

The connectivity  coefficient $\gamma_i$ can be easily defined 
from the boundary condition for the magnetic field $\vec B$ assuming that  
there is no external field, $H_0=0$, 
 and a permanently magnetized layer $j$ is the only source that generates a 
magnetic field in the layer $i$. 
 For $i=1$ and $j=2$, the case shown in Figure~\ref{fig-2},
the field inside layer   {\it 2} is  
\begin{equation}
\vec B_2 = \mu_0 (\vec H_2 + \vec M_2).
\label{eq-3}
\end{equation} 
Outside layer {\it 2}, at its bottom boundary, 
\begin{equation}
\vec B_{out} = \mu_0 \vec H_{out},
\label{eq-4}
\end{equation} 
Putting 
\begin{equation}
\vec H_2 = \vec H_0 - \alpha_2 \vec M_2 = -\alpha_2 \vec M_2 
\label{eq-4a}
\end{equation} 
into  Eq.(\ref{eq-3}), we get
\begin{equation}
\vec B_2 = \mu_0  (1-\alpha_2) \vec M_2.
\label{eq-5}
\end{equation} 
Applying the boundary condition for the continuity of the perpendicular component
 of $\vec B$ at the bottom boundary of layer {\it 2}, $ \vec B_{out} =  \vec B_2$, 
we get from Eq.(\ref{eq-4}) and Eq.(\ref{eq-5})  
\begin{equation}
\vec H_{out} =  (1-\alpha_2) \vec M_2.
\label{eq-6}
\end{equation} 

Within the {\it "upper side"} approach $R_1 = H_{out}$, and  using 
Eq.(\ref{eq-2a}) and  Eq.(\ref{eq-6})  
 we arrive at the {\it preliminary} connectivity coefficient 
\begin{equation}
\widetilde{\gamma_1} =  1-\alpha_2.
\label{eq-7}
\end{equation} 
However, within our generalized approach $R_1<H_{out}$, and thus 
using $ R_1  = \beta_1 H_{out}$,  where $\beta_1<1$ is a coefficient that, generally speaking, depends 
on the coefficients $\alpha_1$ and $\alpha_2$,
we get  for the connectivity coefficient
\begin{equation}
\gamma_1 =  \beta_1 (1-\alpha_2).
\label{eq-7a}
\end{equation} 

The exact value of $\beta_1$ can be calculated only using numerical procedures.  In our analytical approach we can  
define the upper limit for $\beta_1$, above which non-physical effects of negative magnetization might take place, see 
Appendix A for more details. 

In a similar manner we define the connectivity coefficient for the second layer as 
$\gamma_2 =  \beta_2 (1-\alpha_1)$. It is worth to mention that, for an
infinitely wide ($L \gg L_z$) prism $\alpha_i=1$ ($i$=1,2), and 
the connectivity coefficients $\gamma_i=0$ regardless  of the values of 
$\beta_i$, meaning that  $R_i(H_j)=0$. In other words, the cross term $R_i(H_j)$ is negligible for flat geometries.

Finally we arrive at the following relation for the field
 $\vec H_i$ inside the layer $i$ of the prism placed under the external field $\vec H_0$, 
\begin{eqnarray}
\vec H_1  =  \vec H_0 - \alpha_1 \vec M_1 + \beta_1 (1-\alpha_2) \vec M_2, \nonumber \\
\vec H_2  =  \vec H_0 - \alpha_2 \vec M_2 + \beta_2 (1-\alpha_1) \vec M_1. 
 \label{eq-8}
\end{eqnarray} 

Below, for simplicity, we will assume that $\beta_1 = \beta_2 = \beta$ in order to proceed 
to analytical results. Thus putting $\vec H_i = \vec M_i / \chi_i $, where $\chi_i$ is the
susceptibility of the layer $i$ we find 
\begin{eqnarray}
M_1  =  H_0  \frac 
{ \chi_1 (1+ \alpha_2 \chi_2) + \beta \chi_1 \chi_2 (1-\alpha_2) } 
{ (1 + \chi_1 \alpha_1)(1+\chi_2 \alpha_2)  - \beta^2 (1-\alpha_1)(1-\alpha_2) \chi_1 \chi_2}, 
\nonumber \\
M_2  =  H_0  \frac 
{ \chi_2 (1+ \alpha_1 \chi_1) + \beta \chi_1 \chi_2 (1-\alpha_1) } 
{ (1 + \chi_1 \alpha_1)(1+\chi_2 \alpha_2)  - \beta^2 (1-\alpha_1)(1-\alpha_2) \chi_1 \chi_2}. 
 \label{eq-9}
\end{eqnarray} 
 For a single layer, i.e.  when $\chi_2 = 0$, 
from Eq.(\ref{eq-9}) we recover the magnetization $M_1$ of the single layer
\begin{equation}
\vec M_1 = \vec  H_0 \frac{\chi_1}{ 1 + \alpha_1 \chi_1}.
\label{eq-10}
\end{equation}

Eq.(\ref{eq-9}) is the main result for the magnetization of the bilayer and will be used to calculate 
the magnetic pressure on the composite prism  in the next section.

\FloatBarrier    
\clearpage
\newpage

\section{Magnetic pole distribution at the bilayer interface}

\begin{figure}  [!ht]
\begin{center}
\includegraphics*[width=0.8\textwidth,height=0.6\textwidth]{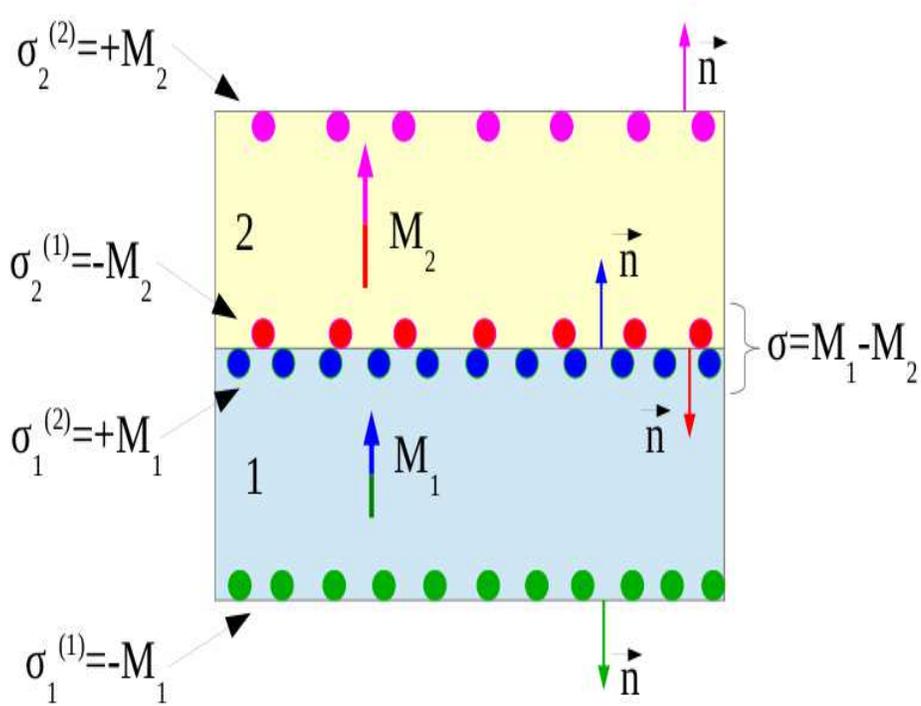}
\end{center}
\caption{(Color in online)  Distribution of the magnetic poles at the prism boundaries. 
The sign of the poles is defined from $\sigma_i^{(j)} = \left(\vec M_i  \cdot \vec n \right)_j$,
where $j$ indicates the bottom ($j=1$) and upper ($j=2$) boundaries of each layer $i=1,2$.
$\vec n$ is a unit vector pointing outward from the layer surface. 
 \label{fig-3}
}
\end{figure}

For the set-up presented in Figure~\ref{fig-1} the distribution of the poles is schematically
shown in Figure~\ref{fig-3}. For the pole density on layer {\it 1}  we have  
$\sigma_1^{(1)} = - M_1$ for the bottom and  
$\sigma_1^{(2)}   = + M_1$ for the top boundaries, and for layer {\it 2}
the density of boundary poles  are $\sigma_2^{(1)} = - M_2$ and  
$\sigma_2^{(1)}   = + M_2$ correspondingly. 
As a result, the net magnetic pole density at the interface between
layers {\it 1} and {\it 2} is  
\begin{equation}
\Delta \sigma = \sigma_1^{(2)} +\sigma_2^{(1)} = M_1 - M_2,
\label{eq-12}
\end{equation} 
 or, taking into account Eq.(\ref{eq-9}),
\begin{equation}
\Delta \sigma =  H_0 
 \frac{\chi_1 - \chi_2  + (\alpha_1 - \alpha_2)(1-\beta) \chi_1 \chi_2}  
{ (1 + \chi_1 \alpha_1)(1+\chi_2 \alpha_2)  - \beta^2 (1-\alpha_1)(1-\alpha_2) \chi_1 \chi_2}. 
\label{eq-13}
\end{equation} 

From Eq.(\ref{single-force-12}) for the driving pressure $p = \frac{F}{S} = \frac{F_1-F_2}{S}$
acting on the interface {\it 1-2}
we have 
\begin{equation}
p =   \mu_0 \vert \vec H_0 \vert 
\left[
A(\alpha_1,\chi_1) \vec M_1  - A(\alpha_2,\chi_2) \vec M_2
\right]  \cdot  \vec n.
\label{eq-13-a}
\end{equation} 

This expression reduces to a simple form
\begin{equation}
p =  \mu_0 \vert \vec H_0 \vert 
 A(\alpha) \Delta \sigma  
\label{eq-14}
\end{equation} 
for  $\alpha_1=\alpha_2=\alpha$, $\chi_1 \ll 1$, and $\chi_2\ll 1$, hence $A(\alpha) = (1-2\alpha)$. 
A positive (negative) $p$ in  Eq.(\ref{eq-13-a}) and  Eq.(\ref{eq-14})  means that the layer 
{\it 1} will be stretched (squeezed) into the layer {\it 2}, whereas the layer {\it 2} will be squeezed (stretched).

As  has been mentioned in section 2, both the thickness $L_z$ and 
the area $S$ of the prism deform under the constant volume condition. The total change 
$\Delta L_z$ of the bilayer thickness  is  a sum of the thickness changes in each layer,   
\begin{equation}
\Delta L_z = \sum_{i=1}^2 \Delta d_i,
\label{eq-14-a}
\end{equation}
where $\Delta d_i$ is defined through Hooke's relation for the boundary  forces $F_1$ and $F_2$, 
\begin{eqnarray}
\, \, \, \, \,  F + F_1 =  S Y_1 \frac{\Delta d_1}{d_1},  \nonumber \\
               -F + F_2 =  S Y_2 \frac{\Delta d_2}{d_2},
\label{eq-14-b}
\end{eqnarray}
where $ F_i$, $i=1,2$ is given by Eq.(\ref{single-force-12}).

Putting everything together we have for the strain $\Sigma_B = \Delta L_z/L_z$, 
\begin{equation}
\Sigma_B =        
\frac{d_1}{L_z} 
\left( 
 \frac{ F +  F_1}{S Y_1} +   \frac{F - F_2}{S Y_2}
    \right) +  
 \frac{  F_2 - F}{S Y_2}.
\label{eq-strain}
\end{equation}

This expression, together with the definitions for the forces $F_i$ 
\begin{equation}
\frac{F_i}{S} = \mu_0 \vert \vec H_0 \vert  A(\alpha_i,\chi_i) (\vec M_i  \cdot  \vec n) 
\label{mp-1}
\end{equation}
and the magnetization $M_i$ defined by Eq.(\ref{eq-9}) constitute our main result for the reaction of the bilayered 
magnetic elastomer to the applied field $H_0$.
Note that whereas the pole distribution term $(\vec M_i   \cdot  \vec n)$
and the geometry factor $ A(\alpha_i,\chi_i)$     
in the magnetic pressure equation Eq.(\ref{mp-1}) together determine the magnetic force on the layer 
{\it i} due to the external field, the total deformation of the layer {\it i}  is regulated by the forces given in Eq.(\ref{eq-14-b})

\newpage
\clearpage
\section{Results}
The strain $\Sigma_B$  in Eq.(\ref{eq-strain}) depends on the forces $F_i$,
the conformation parameter $x=d_1/L_z$, and the elasticity moduli $Y_i$.  
The forces  $F_i$, according to Eq.(\ref{eq-9}) and   Eq.(\ref{mp-1}), are also functions of
the four parameters $\alpha_1$, $\alpha_2$ and $\chi_1$, $\chi_2$:  
\begin{eqnarray}
F_i = F_i(\alpha_1,\chi_1,\alpha_2,\chi_2). \nonumber \\
\end{eqnarray}
In total, the strain $\Sigma_B$ depends on the six parameters making 
the analyses of the strain $\Sigma_B$ a very complicated task. 
However, a consideration of  the relative strain, defined as 
\begin{equation}
\Sigma = \Sigma_B/\Sigma_S,
\label{relstr}
\end{equation}
where   the single layer strain is
\begin{equation}
\Sigma_S = \frac {F_3}{S Y_2}
\label{sls}
\end{equation}
(assuming that the single layer is made of material {\it 2}, has the same thickness 
$L_z$ as the composite prism, 
and $F_3/S$ is the magnetic pressure acting on this single-layered reference sample under an identical external magnetic field),
brings the number of independent system parameters from six down to four. 

\subsection{ Relative strain of the bilayered composite}
The relative strain $\Sigma$ in Eq.(\ref{relstr}) measures how effective the reaction of the bilayered structure to the applied field is, 
and can be rewritten in parametric form as 
\begin{equation}
\Sigma =   (1 - x)\left(2\theta -\frac{f}{2}\right)  + (2f-\theta) \frac{x}{y}.
\label{eq-strain-relative}
\end{equation}
Here  we have adopted  $f = \frac{F_1}{F_3}$, $y=\frac{Y_1}{Y_2}$, $x = \frac{d_1}{L_z}$, and 
$\theta = F_2/F_3$. The strain $\Sigma$ now depends on four variables instead of six variables for 
$\Sigma_B$, which 
makes its analyses relatively simple. We can fix $\theta$ and $x$, and explore the dependence of 
$\Sigma$ on the  parameters $y$ and $f$. 

The most interesting cases are \\
(i) $\Sigma>1$, the case of  strong bilayer stretching (squeezing) relative to the 
single layer stretching (squeezing),  and \\
(ii) $\Sigma<0$, the case of bilayer shrinking (stretching) while 
the single layer stretches (shrinks).

The case $0<\Sigma<1$ corresponds to a weak bilayer reaction and thus is not interesting to us. 
The parameters $f$ and $\theta$ can run between $- \infty$ and 
$+ \infty$, but for simplicity we will restrict 
ourselves to considering  $f>1$ and $\theta>1$,
which corresponds either to the case when $F_1>F_3$, $F_2>F_3$, as well as $F_3>0$, or
to the case
$F_1 < F_3$, $F_2 < F_3$, as well as $F_3<0$.  

From Eq.(\ref{eq-strain-relative}) for $\Sigma>1$ we have the following relation for $y(f)$:
\begin{equation}
y < \frac{x (2 f - \theta)}{1-(1-x)(2\theta-f/2)}.
\label{eq-stretch}
\end{equation} 
Similarly, the condition $\Sigma<0$ leads to the following relation
\begin{equation}
y > \frac{x (2 f -\theta)}{(1-x)(f/2 -2 \theta)}.
\label{eq-shrink}
\end{equation} 

\begin{figure}  [!ht]
\begin{center}
\includegraphics*[width=0.45\textwidth,height=0.45\textwidth]{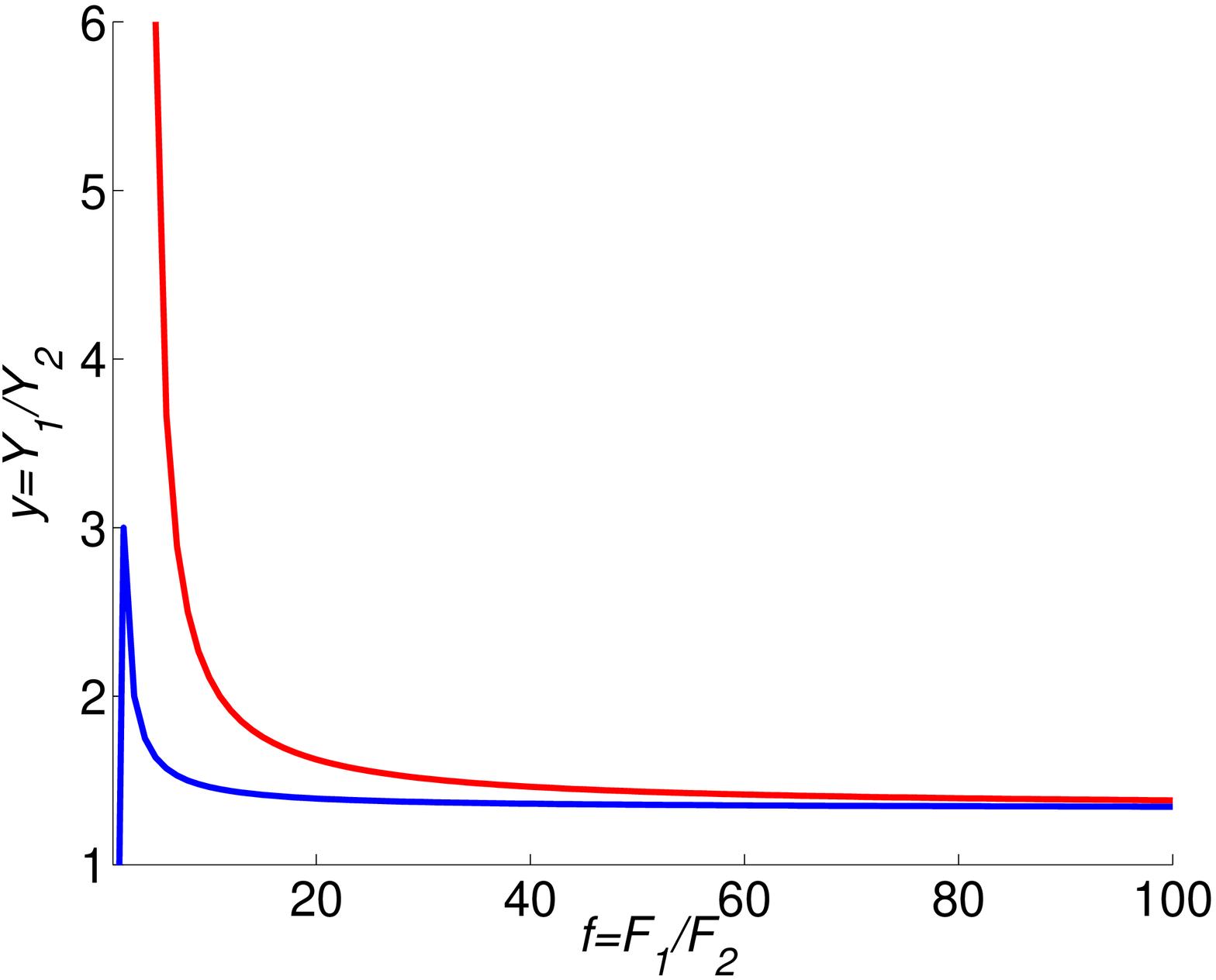}
\includegraphics*[width=0.45\textwidth,height=0.45\textwidth]{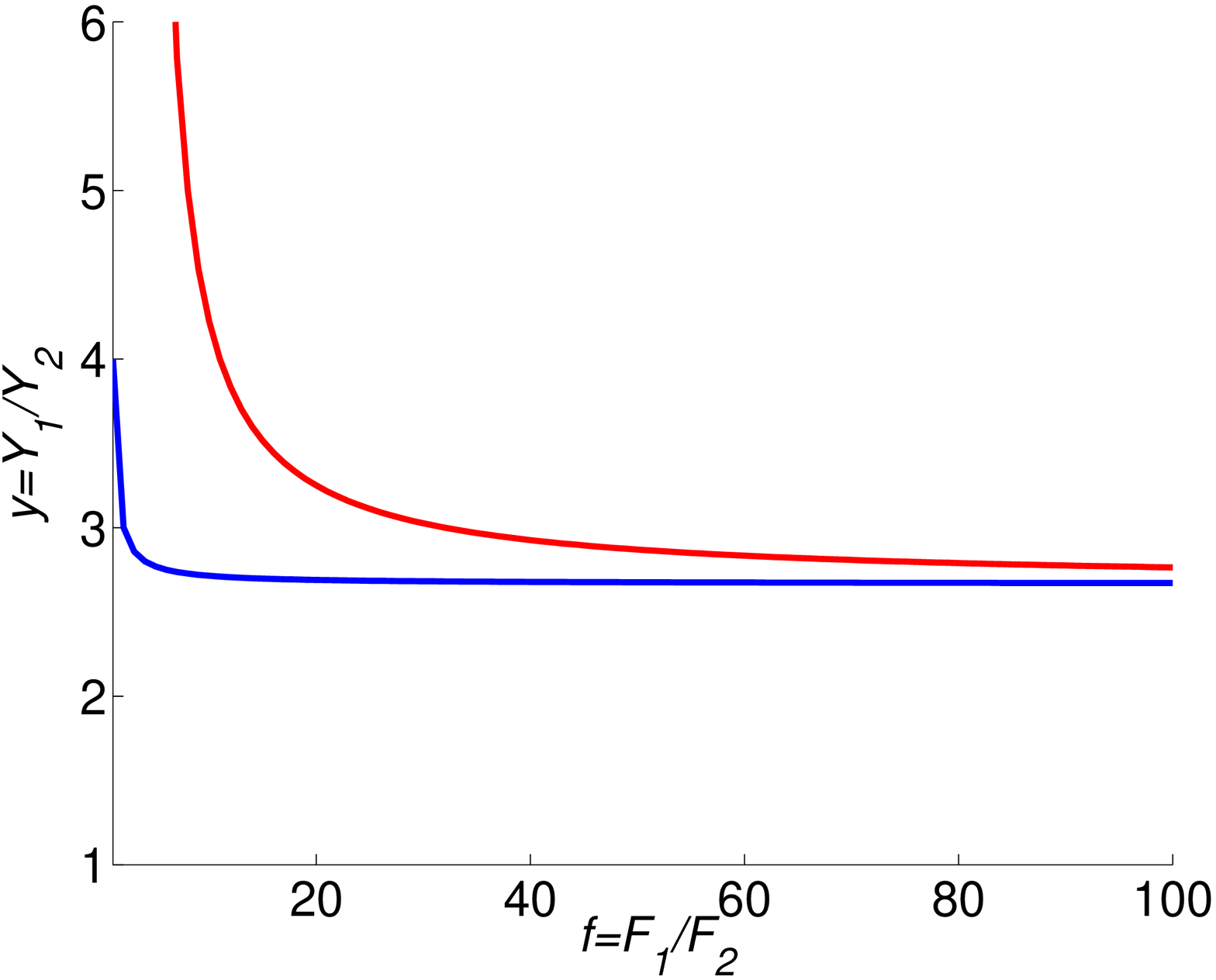}
\includegraphics*[width=0.45\textwidth,height=0.45\textwidth]{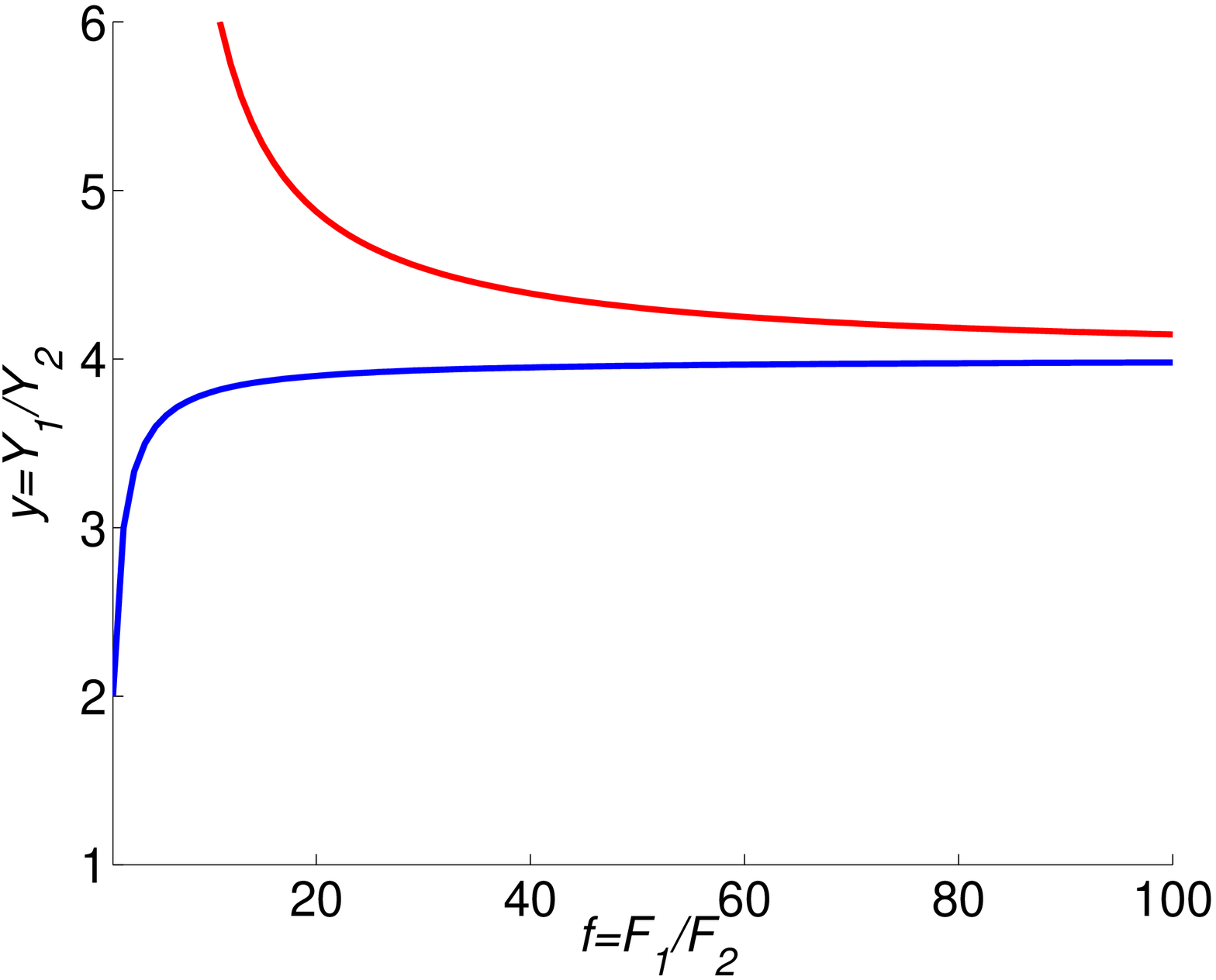}
\includegraphics*[width=0.45\textwidth,height=0.45\textwidth]{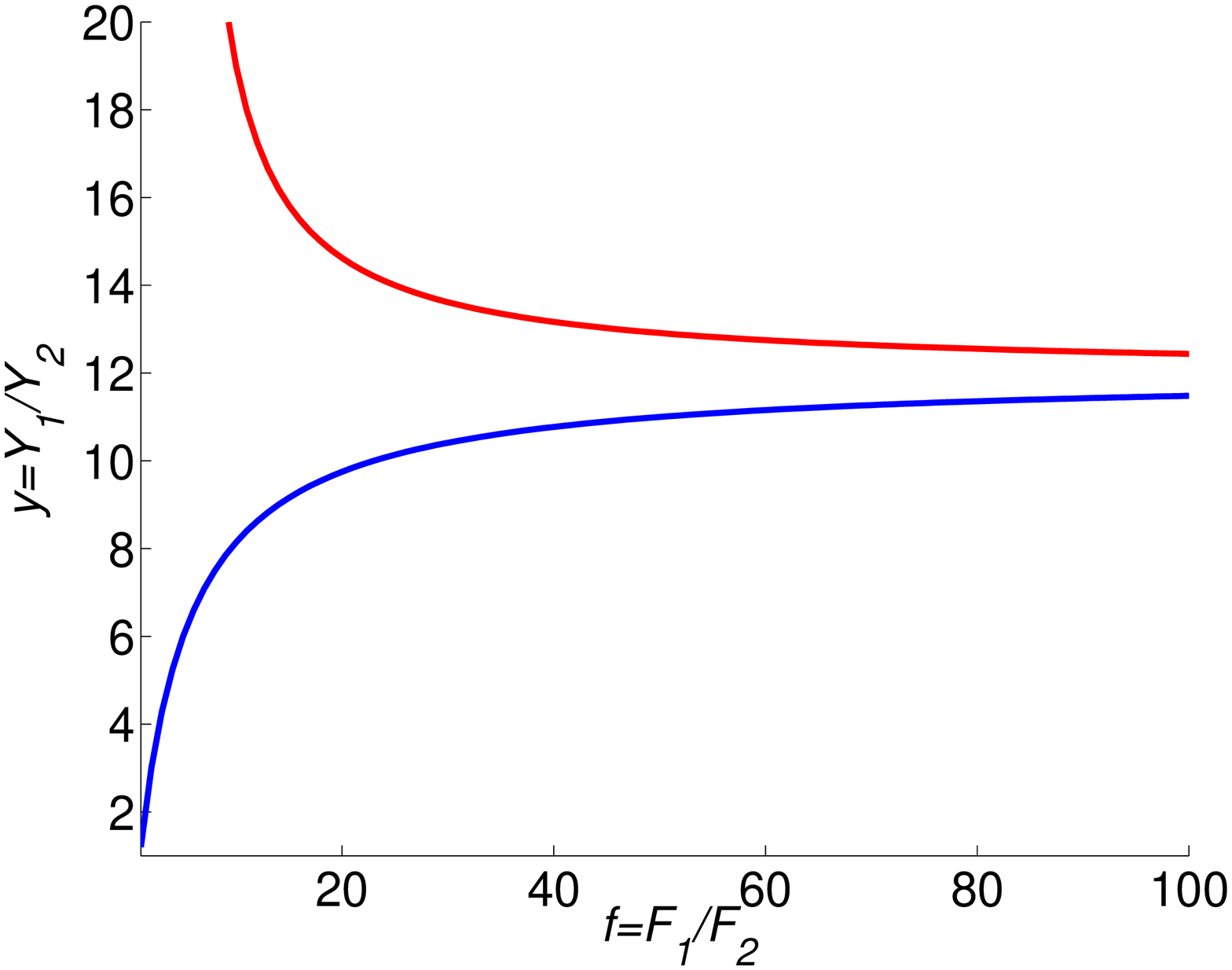}
\end{center}
\caption{(Color in online)  
Logarithmic plot for relative strain curves corresponding to  Eq.(\ref{eq-stretch}) 
and Eq.(\ref{eq-shrink}) for the  composition parameter
 $x = 0.25$ (a), 
 $x = 0.4$ (b),
 $x = 0.5$ (c),
 $x = 0.75$ (d).  
Dashed line (red in color)  corresponds to $\Sigma=0$, solid line (blue in color) is for $\Sigma=1$.
Above the dashed line $\Sigma <0$, and below the solid line $\Sigma>1$. 
 In the  area between these two lines 
$ 0<\Sigma<1$.  
 \label{fig-2D}
}
\end{figure}

Representative pictures for both of these curves, Eq.(\ref{eq-stretch}) 
and Eq.(\ref{eq-shrink}), and for $\theta=1$  are shown in Figure~\ref{fig-2D}.  
It is evident that as the composition factor $x$ increases, a transition from strong stretching ($\Sigma>1$)
to squeezing ($\Sigma<0$) appears at high $y$ values. 
Also, the area of the weak reaction, $0<\Sigma <1$,
widens as the composition factor $x$ increases.

\begin{figure}  [!ht]
\begin{center}
\includegraphics*[width=0.47\textwidth,height=0.4\textwidth]{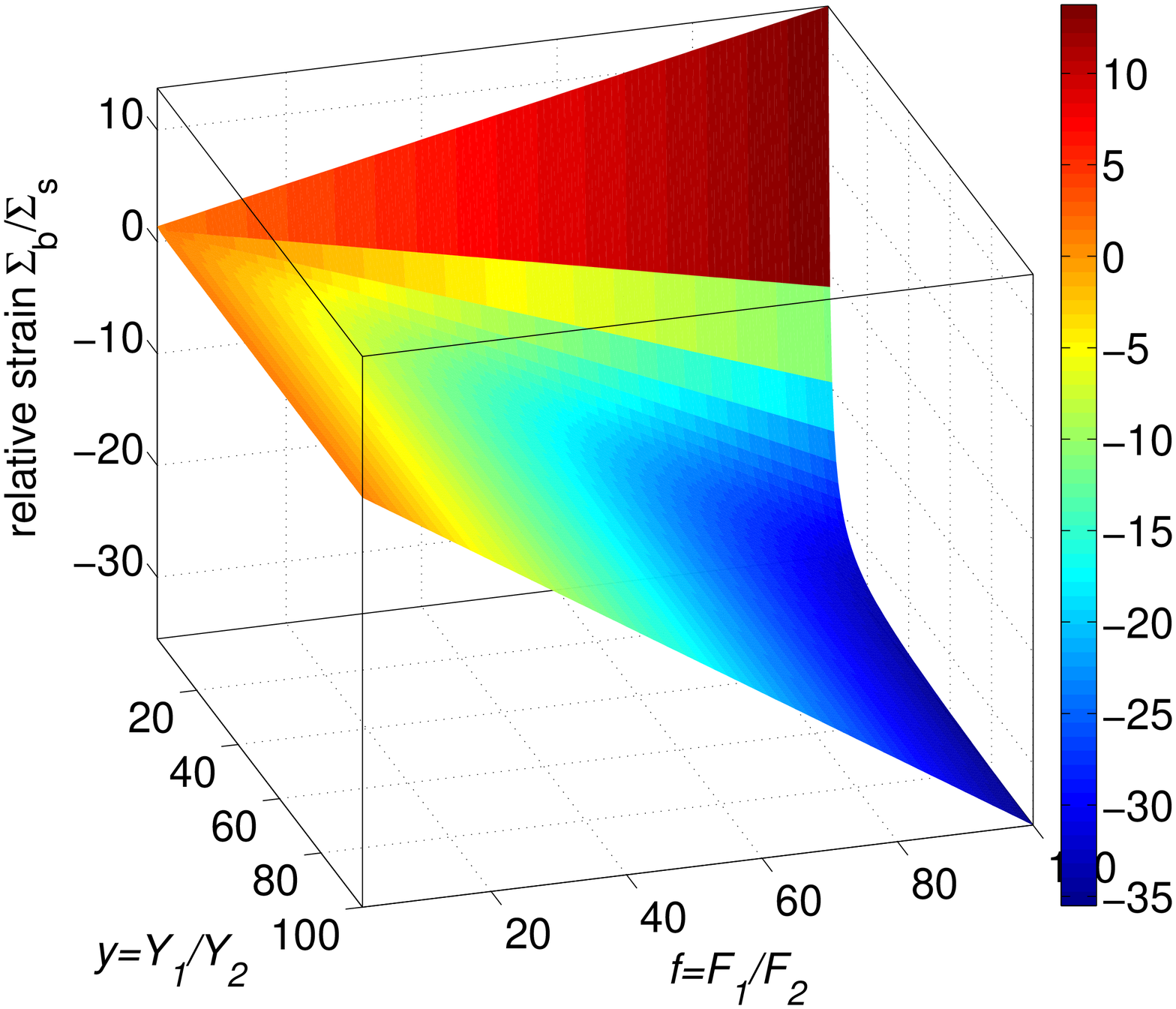}
\includegraphics*[width=0.47\textwidth,height=0.4\textwidth]{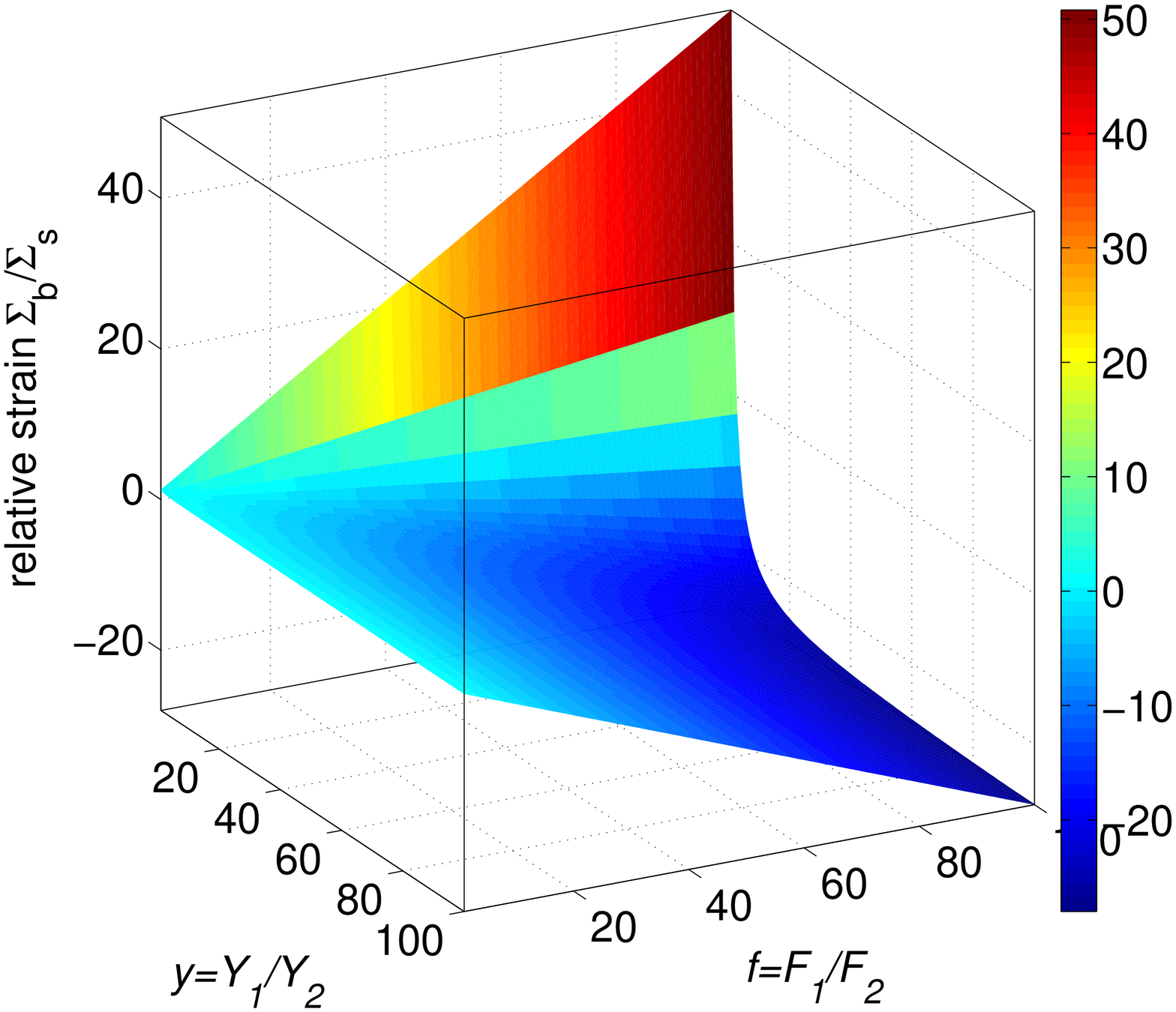}
\includegraphics*[width=0.47\textwidth,height=0.4\textwidth]{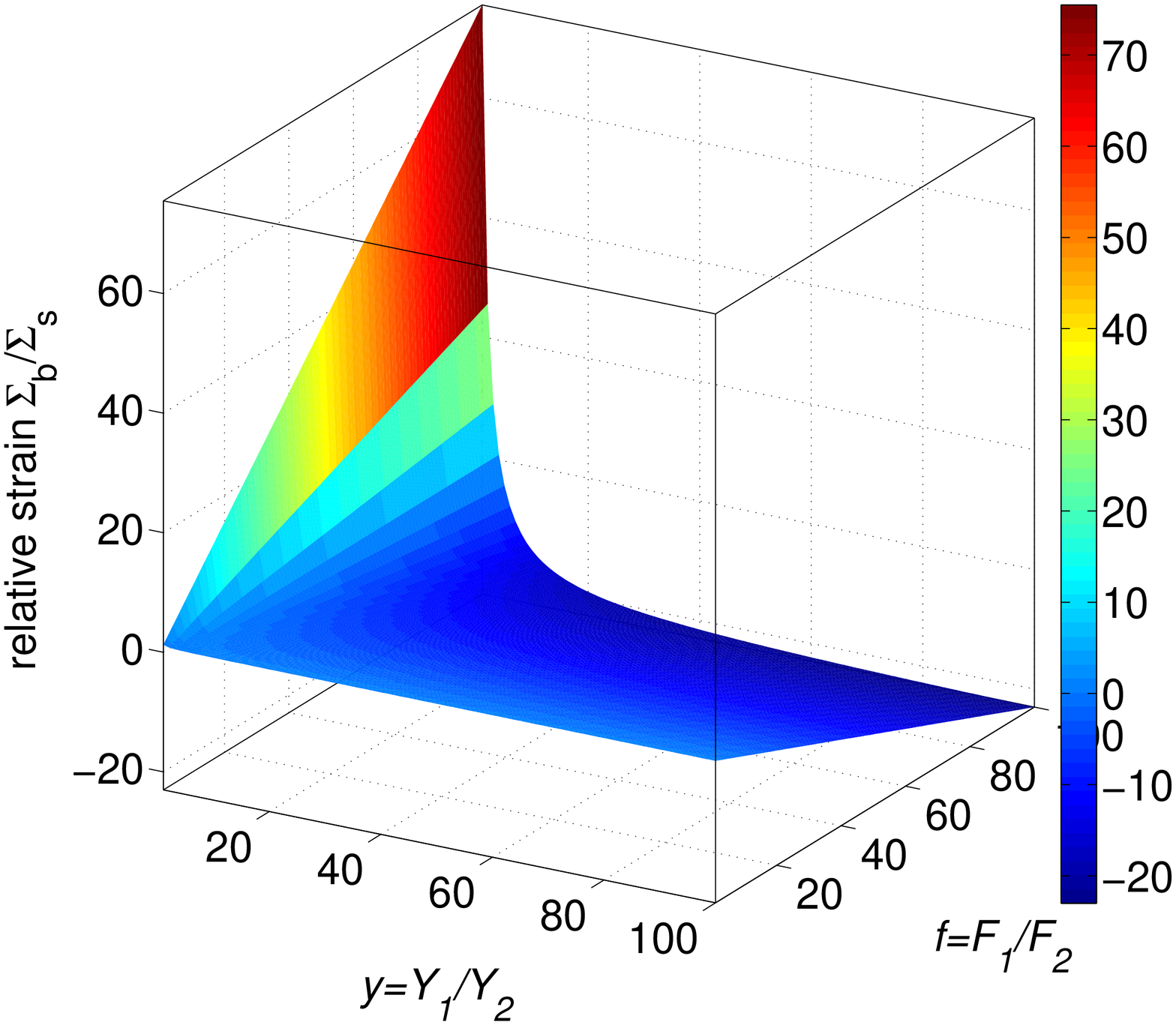}
\includegraphics*[width=0.47\textwidth,height=0.4\textwidth]{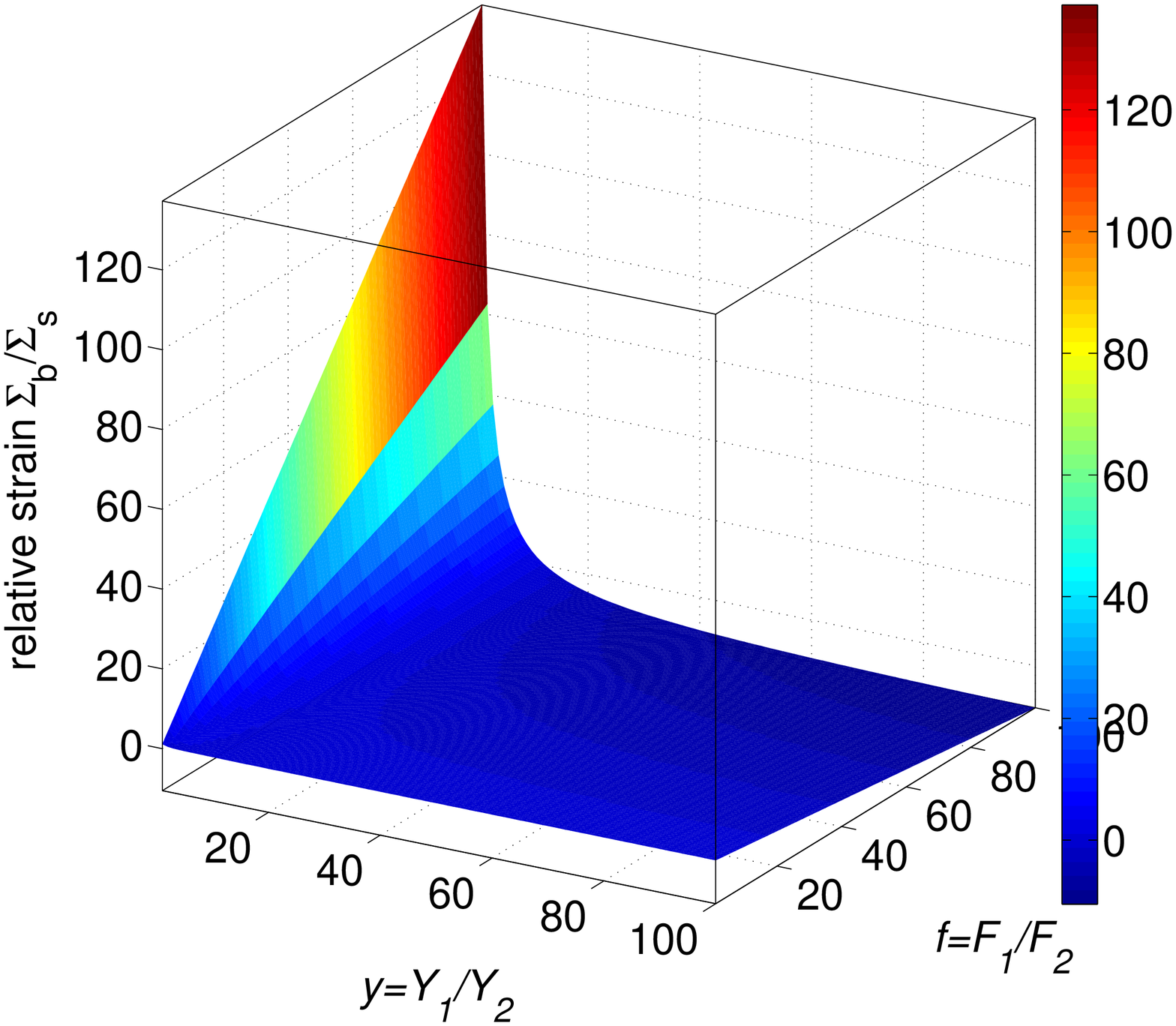}
\end{center}
\caption{(Color in online)  
3D pictures for the relative strain $\Sigma$ as a function of the parameters $y$ and $f$ from 
Eq.(\ref{eq-strain-relative}) and for the composition parameter 
 $x = 0.25$ (a), 
 $x = 0.4$ (b),
 $x = 0.5$ (c),
 $x = 0.75$ (d).  
The color code from dark red  to dark blue corresponds to a decreasing strain strength.
 \label{fig-3D}
}
\end{figure}

The 3D pictures for the relative strain, plotted in Figure~\ref{fig-3D}, show that 
a mild stretching and a strong squeezing at low $x$ is replaced by the  strong stretching and the weak squeezing at large $x$. 

\clearpage
\newpage

\subsection{Full strain of the bilayer composite}

\begin{figure}  [!ht]
\begin{center}
\includegraphics*[width=0.30\textwidth,height=0.30\textwidth]{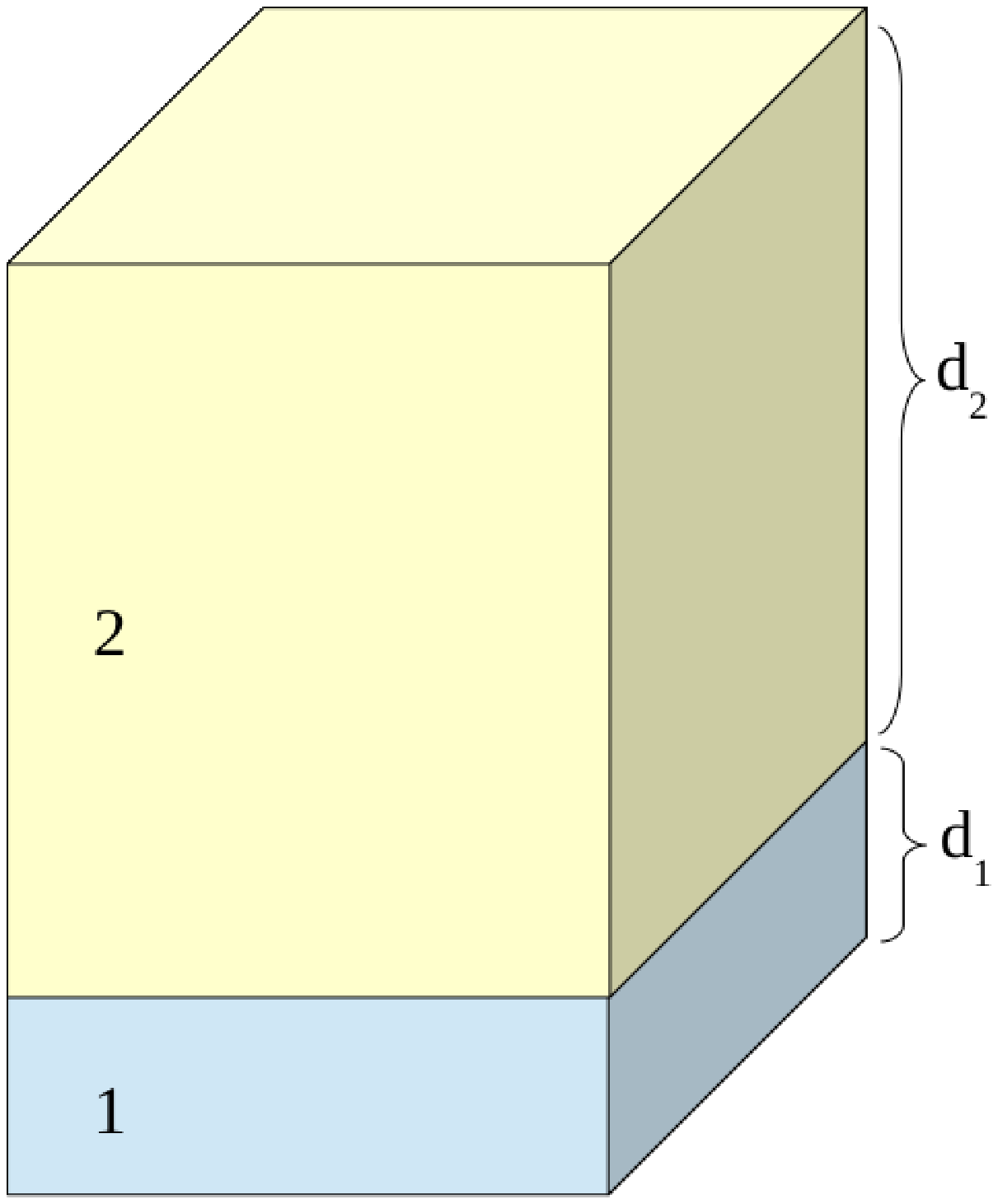}
\includegraphics*[width=0.30\textwidth,height=0.32\textwidth]{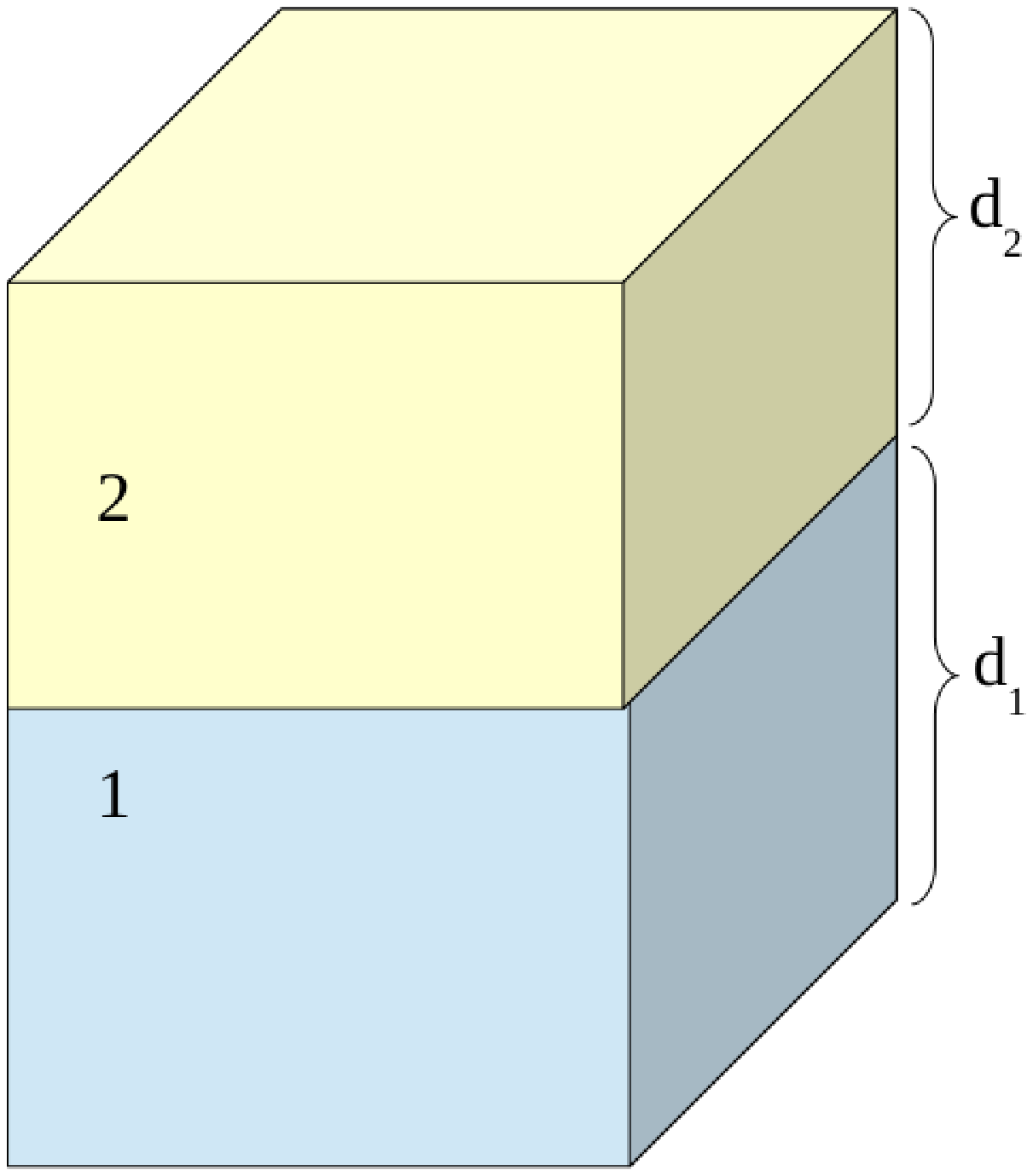}
\includegraphics*[width=0.30\textwidth,height=0.32\textwidth]{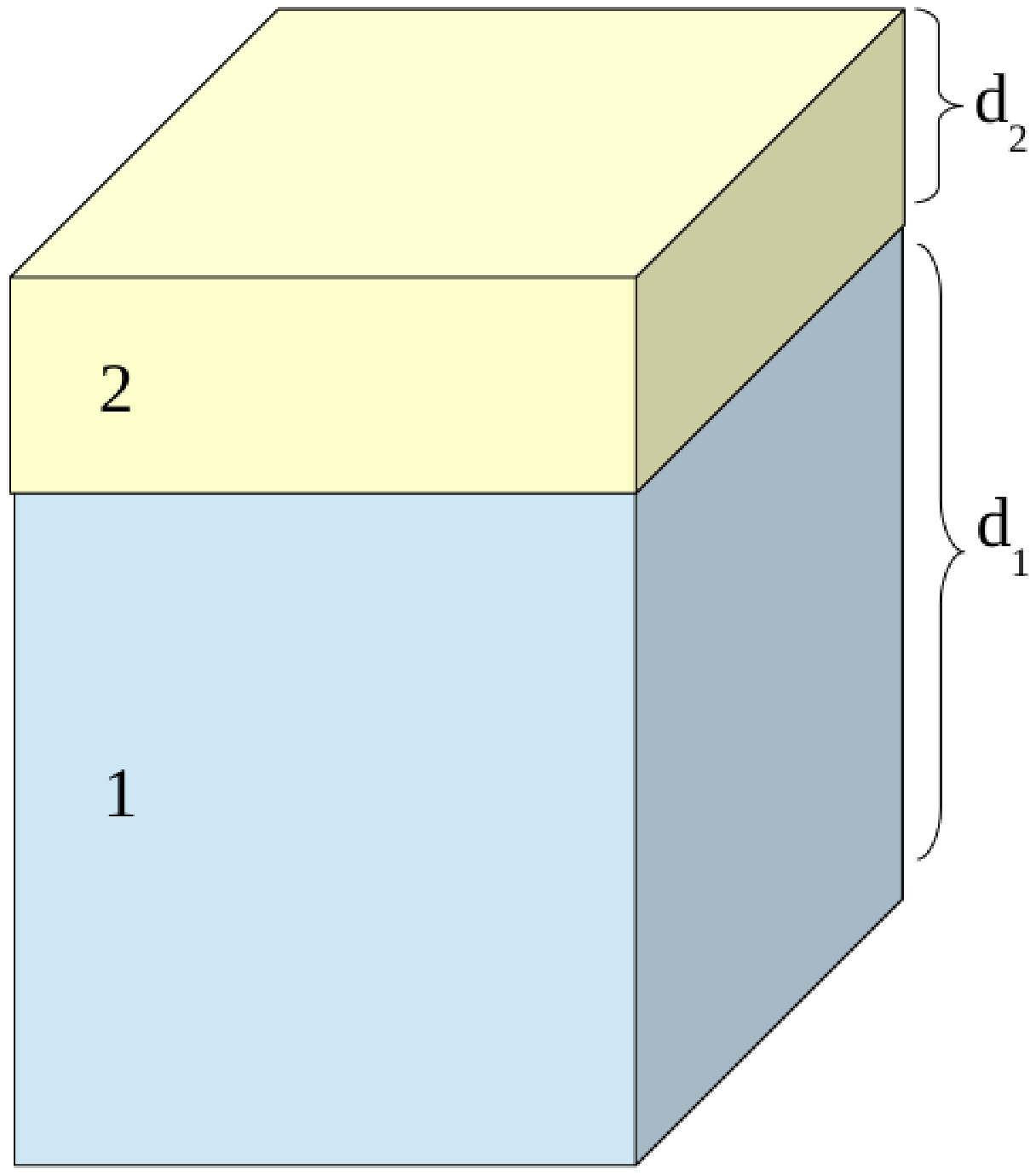}
\includegraphics*[width=0.28\textwidth,height=0.20\textwidth]{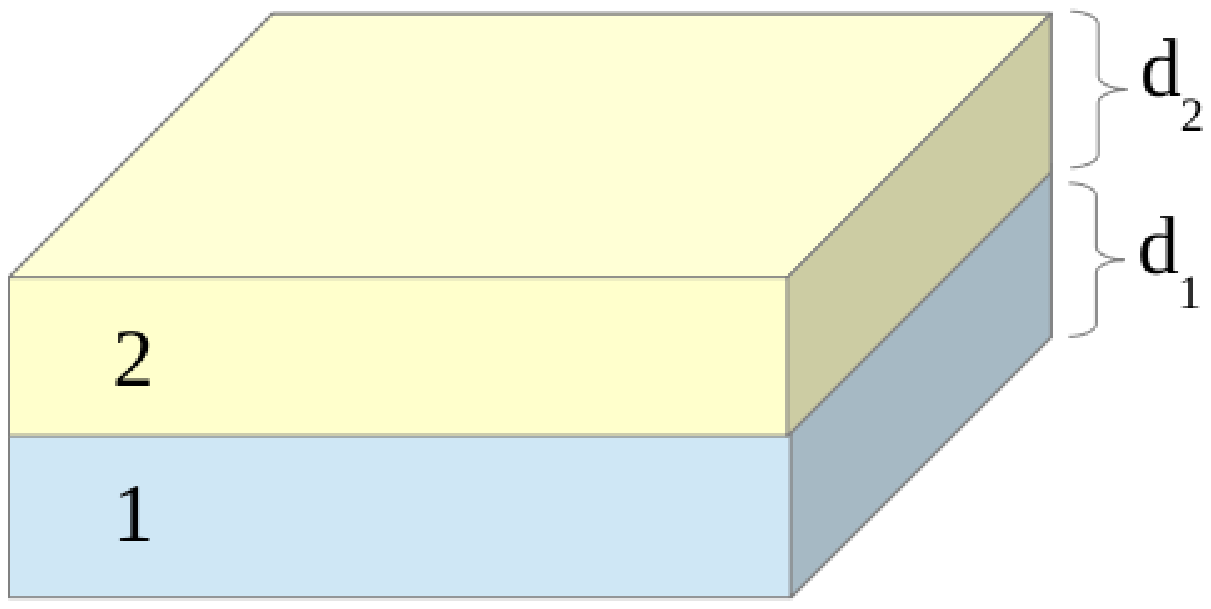}
\end{center}
\caption{(Color in online) 
Geometrical illustration for the 4 different setups from Table~\ref{table-1}. Upper row:  Setups 1, 2, and 3; 
bottom row: Setup 4. 
 \label{fig-4}
}
\end{figure}

In this section we analyze the full strain of the bilayer $\Sigma_B$ given by Eq.(\ref{eq-strain}). 
We  consider three representative cases for $x$, namely $x=0.25,\, 0.5, \, 0.75$. 
Four different setup configurations with the corresponding parameters
 $\alpha_i$ and $A(\alpha_i,\chi_i)$ are shown in Table~\ref{table-1}.  
These setups cover the cases when  the coefficients $A(\alpha_i,\chi_i)$ are
simultaneously either  positive or negative, or have  opposite signs.  Corresponding setup configurations  are 
graphically presented in Figure~\ref{fig-4}. 

\begin{table}[!ht]
\caption{\label{table-1} Geometry-defined demagnetization coefficients $\alpha_i$ and the geometry functions $A(\alpha_i,\chi_i)$
for the three composition parameters  $x$ describing the four different setups in Figure~\ref{fig-4}.
$\chi_1= 1$ and $\chi_2=10^{-3}$ were used to calculate $A(\alpha_i,\chi_i)$.}
\begin{tabular}[t]{| l | c | c | c | c | c |}
\hline
Setup & $x$ & $\alpha_1$ & $\alpha_2$ &  $A(\alpha_1,\chi_1)$  & $A(\alpha_2,\chi_2)$  \\ 
\hline 
\hline 
 1  & 0.25     &   $\frac{2}{3}$    &  $\frac{1}{10}$  & -0.47    &  \, 0.8 \\
\hline 
 2  & 0.5      &   $\frac{1}{3}$    &  $\frac{1}{3}$   & \, 0.25    &  \, 0.33 \\
\hline 
 3  & 0.75     &   $\frac{1}{10}$   &  $\frac{2}{3}$   &  \, 0.72     & -0.33 \\
\hline 
 4  & 0.5      &   $\frac{2}{3}$    &  $\frac{2}{3}$   & -0.47    & -0.33 \\
\hline
\end{tabular}
\end{table}

\begin{figure}  [!ht]
\begin{center}
\includegraphics*[width=0.4\textwidth,height=0.33\textwidth]{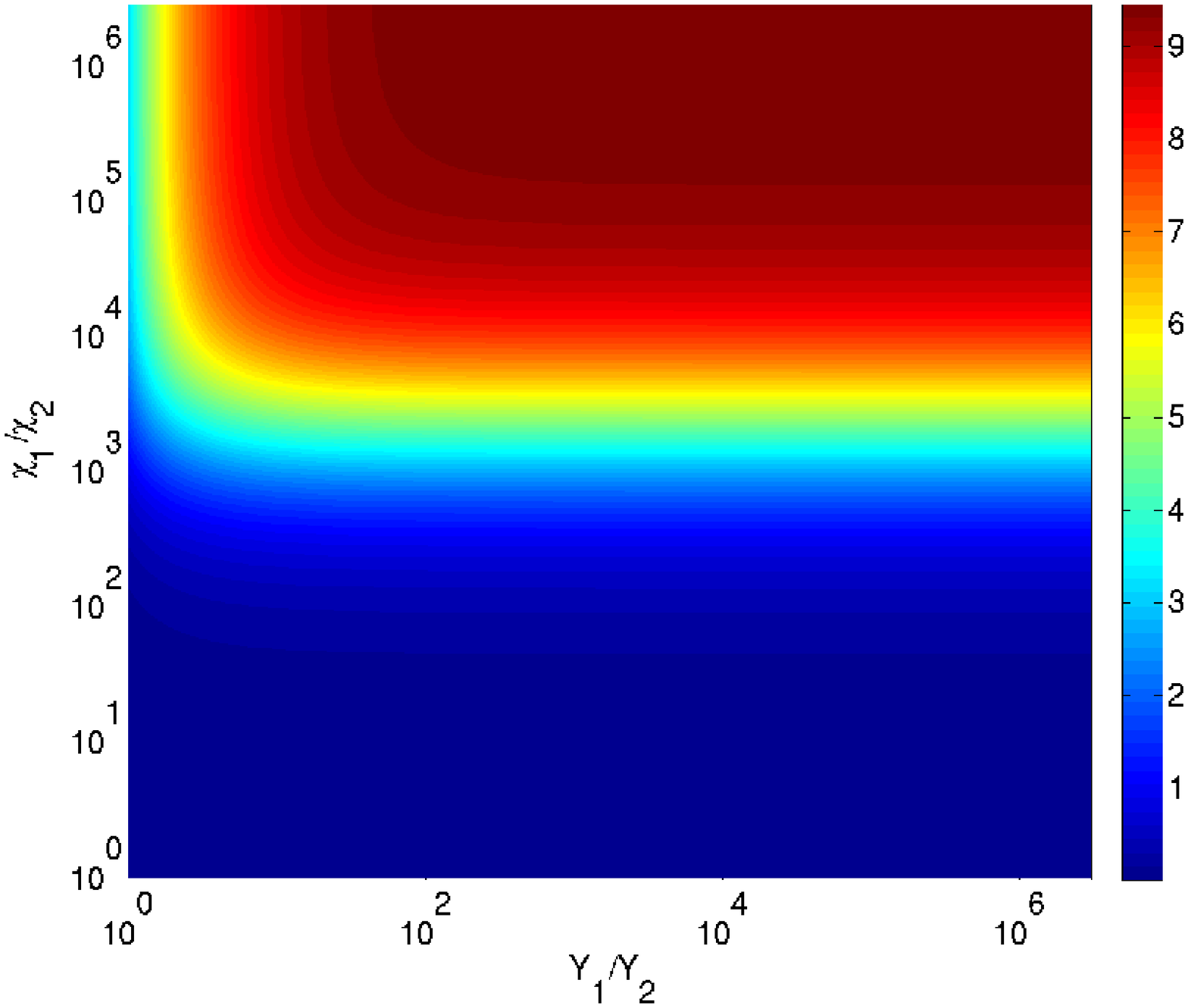}
\includegraphics*[width=0.4\textwidth,height=0.33\textwidth]{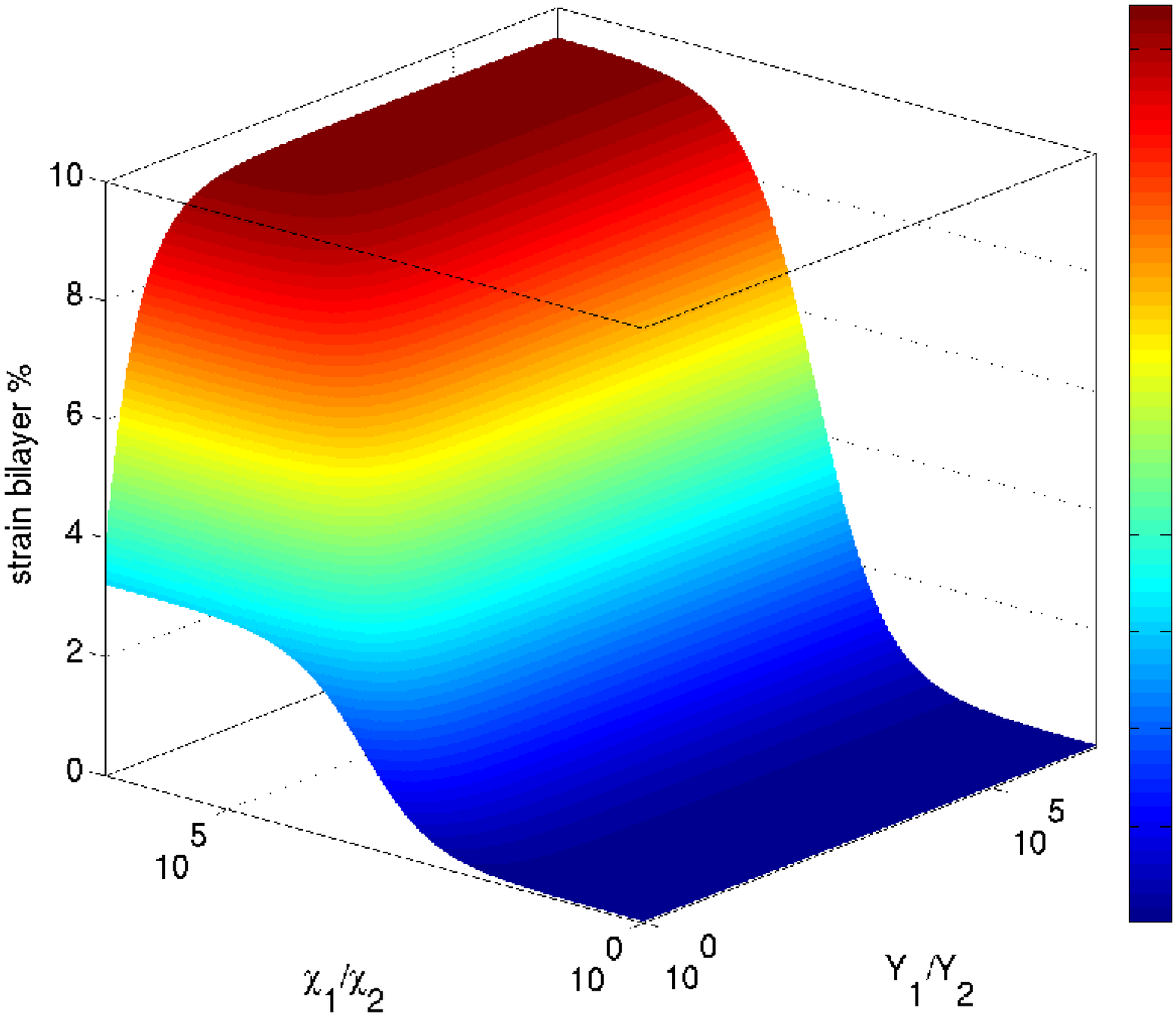}
\includegraphics*[width=0.4\textwidth,height=0.33\textwidth]{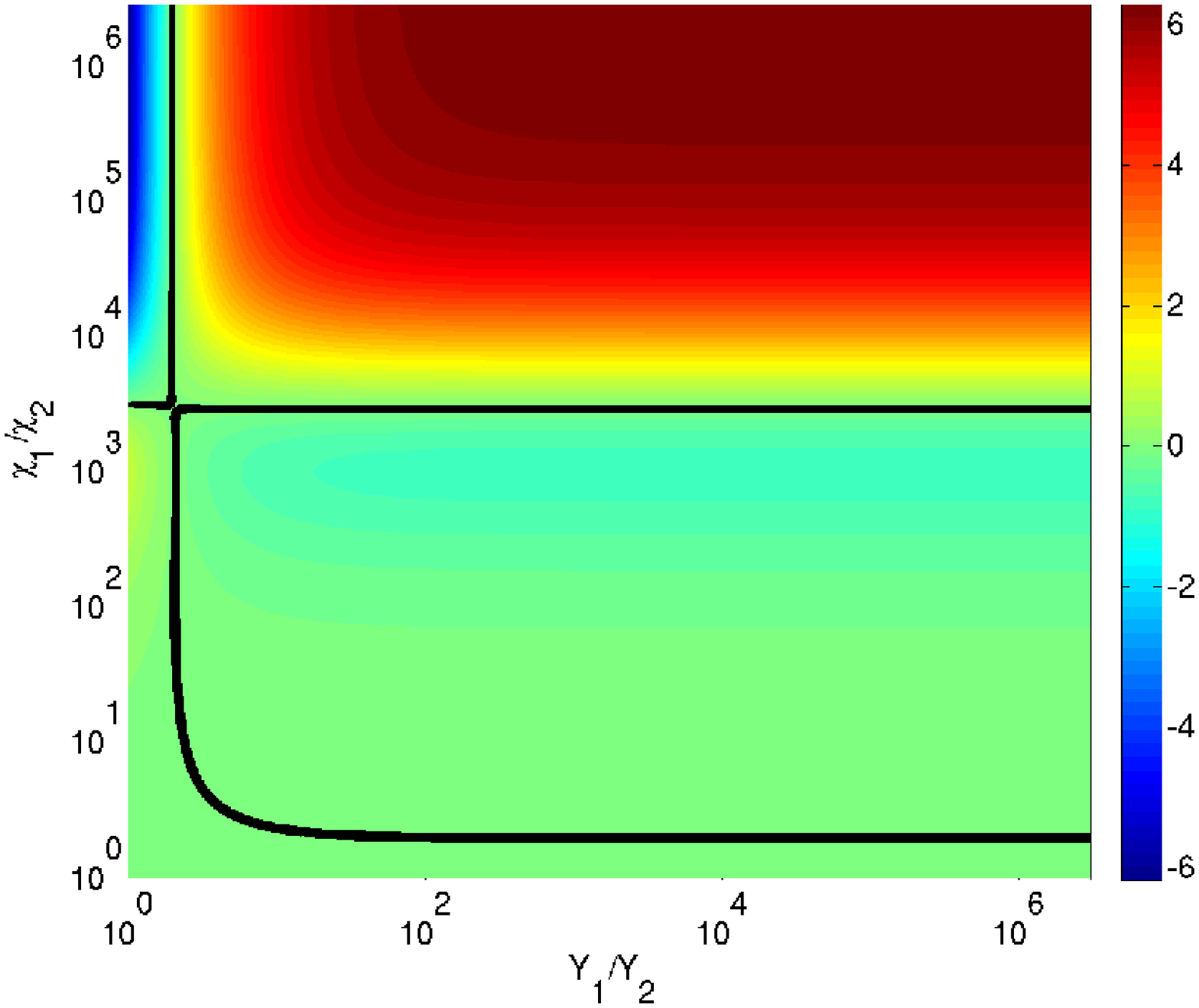}
\includegraphics*[width=0.4\textwidth,height=0.33\textwidth]{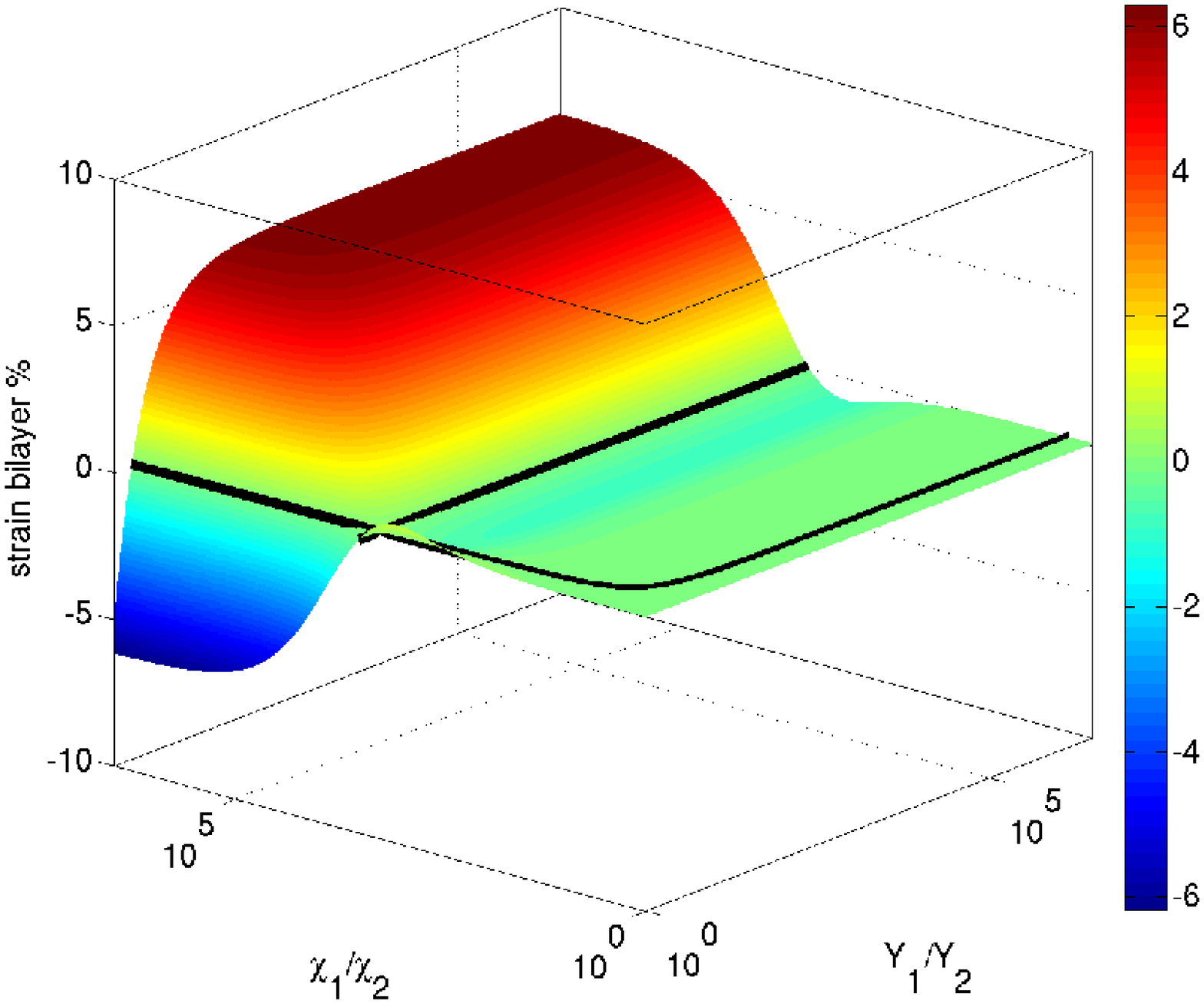}
\includegraphics*[width=0.4\textwidth,height=0.33\textwidth]{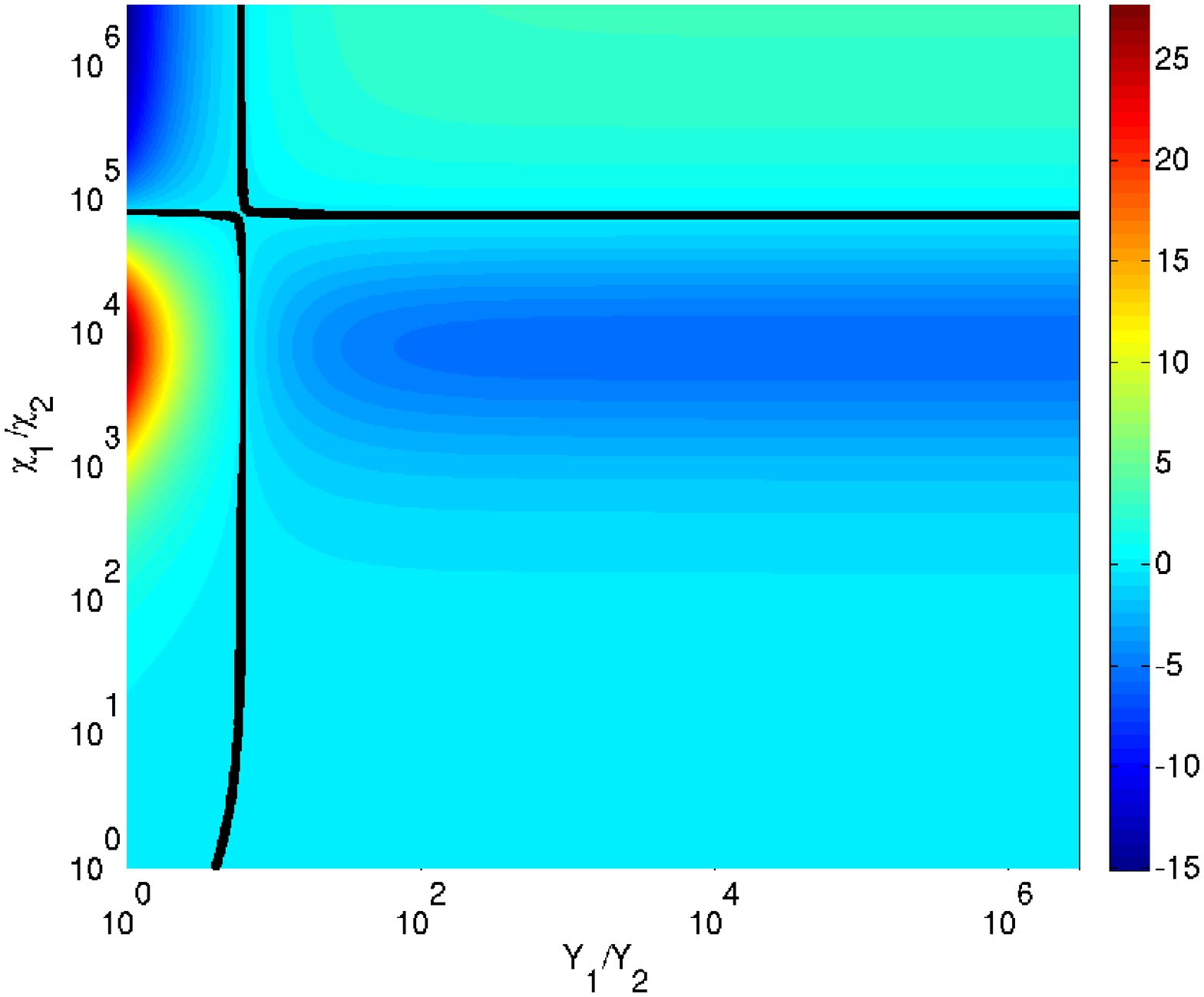}
\includegraphics*[width=0.4\textwidth,height=0.33\textwidth]{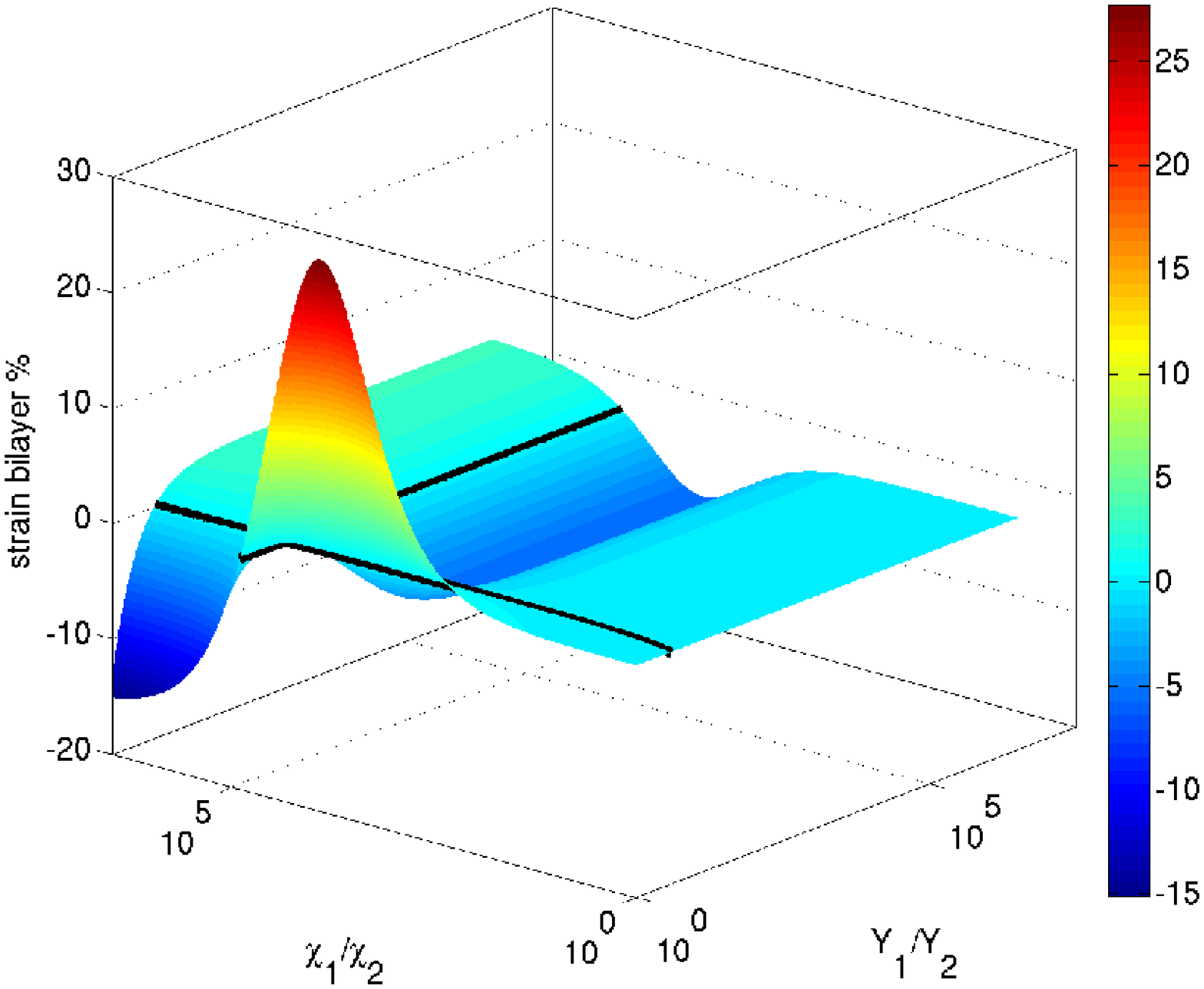}
\includegraphics*[width=0.4\textwidth,height=0.33\textwidth]{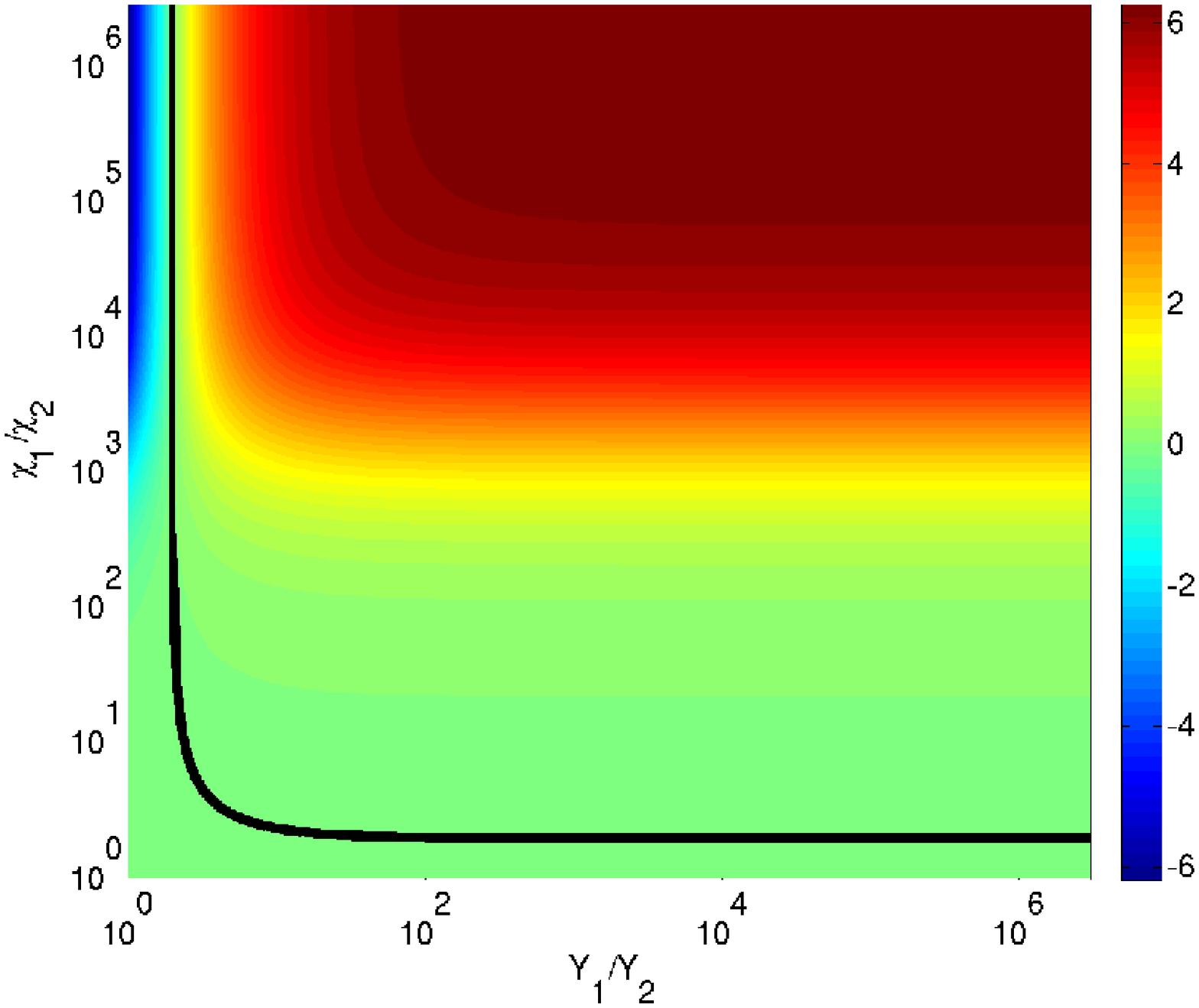}
\includegraphics*[width=0.4\textwidth,height=0.33\textwidth]{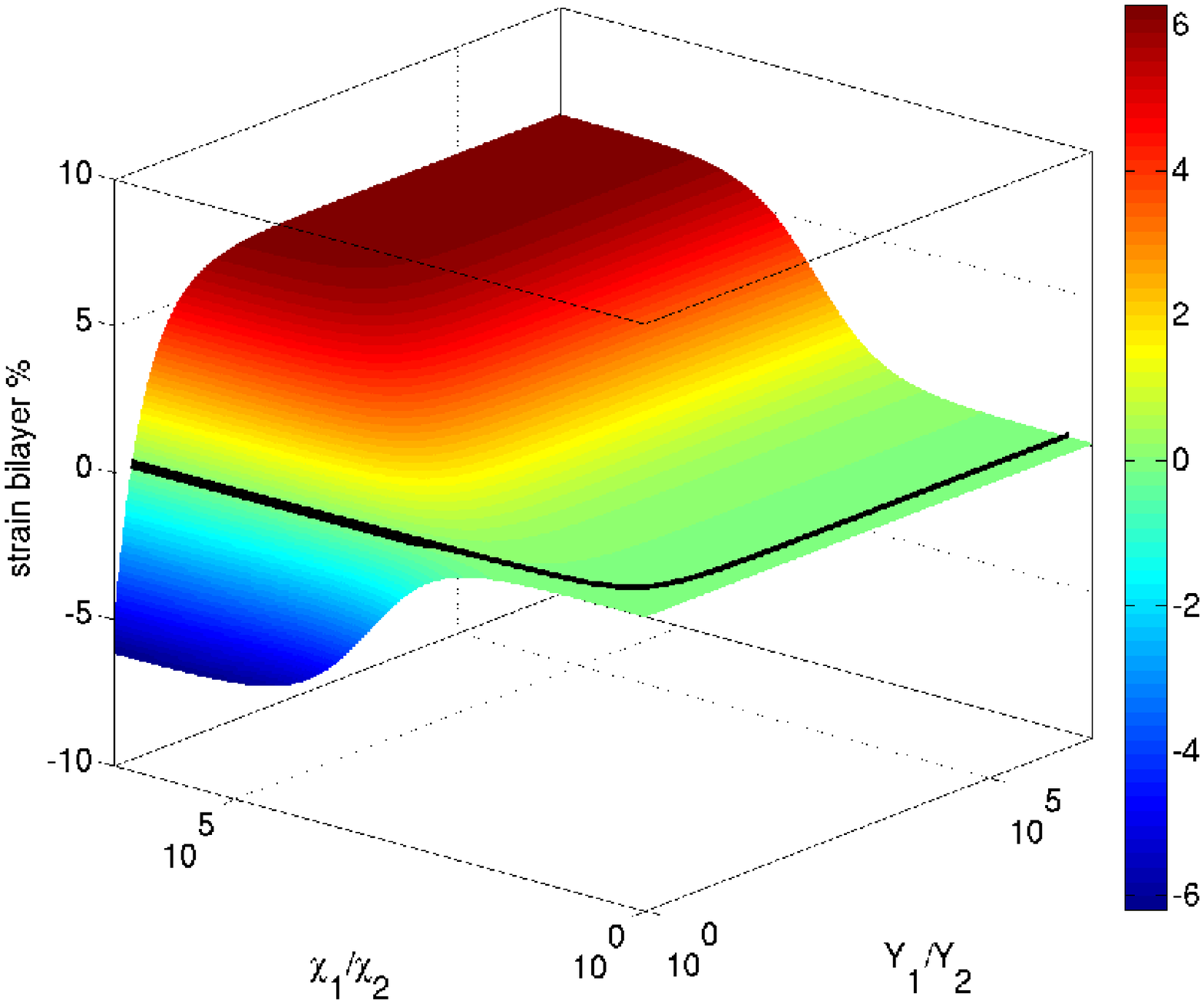}
\end{center}
\caption{(Color in online) 
Logarithmic plot for the bilayer strain from the four Setups given in Table 1. 
From top to bottom, Setup 1 (a), Setup 2 (b), Setup 3 (c), and Setup 4 (d). 
The other system parameters are: 
$L=100 \mu m$, $L_z=200\mu m$, $\chi_2 = 10^{-3}$, $Y_2=10^5 \frac{N}{m^2}$, $B_0=\mu_0 H_0 = 0.13$ Tesla.  
The left picture corresponds to the top view of the 3D surface that is shown on the right. 
Black lines indicate a zero strain of the composite, $\Sigma_B=0$. 
The color code from dark red  to dark blue corresponds to a decreasing strain strength.
 \label{fig-setup-1}
}
\end{figure}

The bilayer strain for Setup 1 as a function of  parameters 
$\chi= \chi_1/\chi_2$ and $y = Y_1/Y_2$ is plotted in Figure~{\ref{fig-setup-1}}a. 
For this case the bilayer has a completely  positive deformation, meaning that it
always experiences a stretching. A relatively high deformation at fixed $y$ 
happens at larger values of $\chi$. 
If an imaginary line at fixed $y=10^4$
is followed from $\chi=1$ to $\chi=10^6$, the composite strain will increase gradually from zero to several percents 
achieving a value of about 10~\% at $\chi>10^5$.

A completely different scenario is observed for Setup 2 with $x=0.5$, see Figure~{\ref{fig-setup-1}}b. In this 
equivalent case when the layers have the same thicknesses $d_1 = d_2$, 
the strain shows both negative and 
positive domains.
The black line corresponds to $\Sigma_B=0$, a zero deformation of the composite for $\Delta L_z$=0. This zero strain 
happens when the changes in the layer $1$ and layer $2$ thicknesses compensate each-other, $\Delta L_1 = -\Delta L_2$.
 A negative strain, or a shrinking of the bilayer along the $z$-axis, takes place at high $\chi$
and low $y$ values. Another negative strain region is 
visible for $\chi< 10^3$ and at about $y>2$. Also, in addition to the strong stretching similar to the Setup 1, 
there is the second, though very mild, stretching in the very tiny strip at low $\chi$ and the $y$ stripes around the bottom left corner of 
the left plot in Figure~{\ref{fig-setup-1}}b. If an imaginary line at fixed $y=10^4$
is followed from $\chi=1$ to $\chi=10^6$, the composite deformation will be first positive, then negative, and then positive again. 
Thus the positive deformation of the composite is reentrant as a function of $\chi$. 

In Setup 3 we again observe two positive and two negative deformation domains, see Figure~{\ref{fig-setup-1}}c.
However, the overall picture is totally different from 
the results for Setups 1 and 2. First, the areas of strong stretching for previous setups now show a small stretching less 
than a few percents. Second, the shrinking of the composite increases, reaching $-15$~\%,
 wheres in Setup 2 it was around $-6$~\%. Third, a visible negative well develops for $10^2<\chi<10^5$. And fourth, 
a strong stretching is visible at very small $y$ around $\chi \approx 10^4$.       
If we again follow an imaginary line at fixed $y=10^4$
and from $\chi=1$ to $\chi=10^6$, the composite deformation will first be negative, then becomes more negative, 
 and then positive. 

Setup 4 has the same composition factor $x=0.5$ as the Setup 2. The only difference between these Setups is the fact that 
in the former case both geometry functions are negative, while in the latter case they are positive. 
As seen from Figure~{\ref{fig-setup-1}}d, here we only have a single positive and a single negative strain domain.
Basically the strain maximum and minimum values stay the same as for Setup 2, but now the 
low $\chi$ stripe at the left bottom corner of Figure ~{\ref{fig-setup-1}}d is negative.
 Again, if an imaginary line at fixed $y=10^4$
is followed from $\chi=1$ to $\chi=10^6$, the composite deformation will first be negative and then positive.

\section{Conclusions}


As explained in the introduction, there are different sources of magnetomechanical coupling in ferrogels and magnetic elastomers. 
The most obvious one is associated with the magnetic interactions between embedded magnetic particles, which can induce mechanical deformations \cite{ivaneyko2011magneto,wood2011modeling,camp2011effects,stolbov2011modelling, ivaneyko2012effects,weeber2012deformation,gong2012full,zubarev2012theory,zubarev2013effect, zubarev2013magnetodeformation,ivaneyko2014mechanical}. 
Furthermore, the aligning magnetic torque onto embedded ferromagnetic particles can directly induce distortions when the particles are chemically crosslinked into the polymer mesh \cite{frickel2011magneto,messing2011cobalt,weeber2012deformation}.

In this paper, we have analyzed a completely different source of magnetomechanical coupling. It results from 
the structural arrangement of two magnetic elastomers into a bilayered composite material. More technically speaking, it follows from the interplay of the magnetic pressures acting on the outer boundaries of the sample and on the internal interfacial boundaries between the layers. 

Using linear response theory for the magnetization and demagnetization fields of a composite 
material of a rectangular prism geometry, we have defined the strain of the bilayer structure to the applied field. 
We have connected the ultimate deformation of the sample to the magnetic pole distribution on the outer 
 boundaries and at the bilayer interface. The material properties of the composite particle, such as its 
susceptibilities and demagnetization coefficients, define a crucial parameter, called 
the geometrical function $A$, which plays a major role in the reaction to 
the applied field. According to our results, the  composite magnetic elastomer is able to
respond more efficiently  to the external field in comparison to a single-component material.
This response also strongly depends on the composition factor of the sample. By changing the composition factor
$x= d_1/L_z$ of the bilayer, it is possible to shift from a mostly stretching composite to a mostly squeezing one when 
all other material parameters are kept fixed.  

Our results are important for the design of optimized bilayered composites of magnetic elastomers and gels. We hope that our analysis will stimulate further research in this direction, both experimentally and theoretically. 
Nevertheless, we are already thinking one step further in a structural hierarchy of magnetic elastomers. Just like magnetic particles embedded in a surrounding polymer matrix in magnetic elastomers or ferrogels, we intend to consider on an upper hierarchical level units of bilayered magnetic elastomers embedded in yet another non-magnetic polymeric matrix. 

Obviously, when the bilayered units stretch along an external magnetic field, the overall hierarchical 
material will elongate along the applied field and get squeezed perpendicular to it due to volume conservation.
In the opposite case, when the bilayered units squeeze along the field direction, 
the overall sample will extend perpendicularly to the field, and its shrinking will be along the field. 
The right management of differently shaped or differently composed bilayered units and 
the right regulation of their embedding places in the overall sample can adjust its overall deformation to the needed demand.   
For example,  it is possible to heterogeneously tune the response of the system during synthesis, 
making it elongate in one part and at the same time shrink in another part.  
All these effects are potentially interesting for their application in a new generation of sensors and in creating new smart (intelligent) materials. 


\section*{Acknowledgments}

A.M.M.\ and H.L.\ thank the Deutsche Forschungsgemeinschaft for support of the work through the SPP 1681 on magnetic hybrid materials.

\clearpage
\newpage

\appendix
\section{Field correction coefficient $\beta$}

\begin{figure}  [!ht]
\begin{center}
\includegraphics*[width=0.5\textwidth,height=0.5\textwidth]{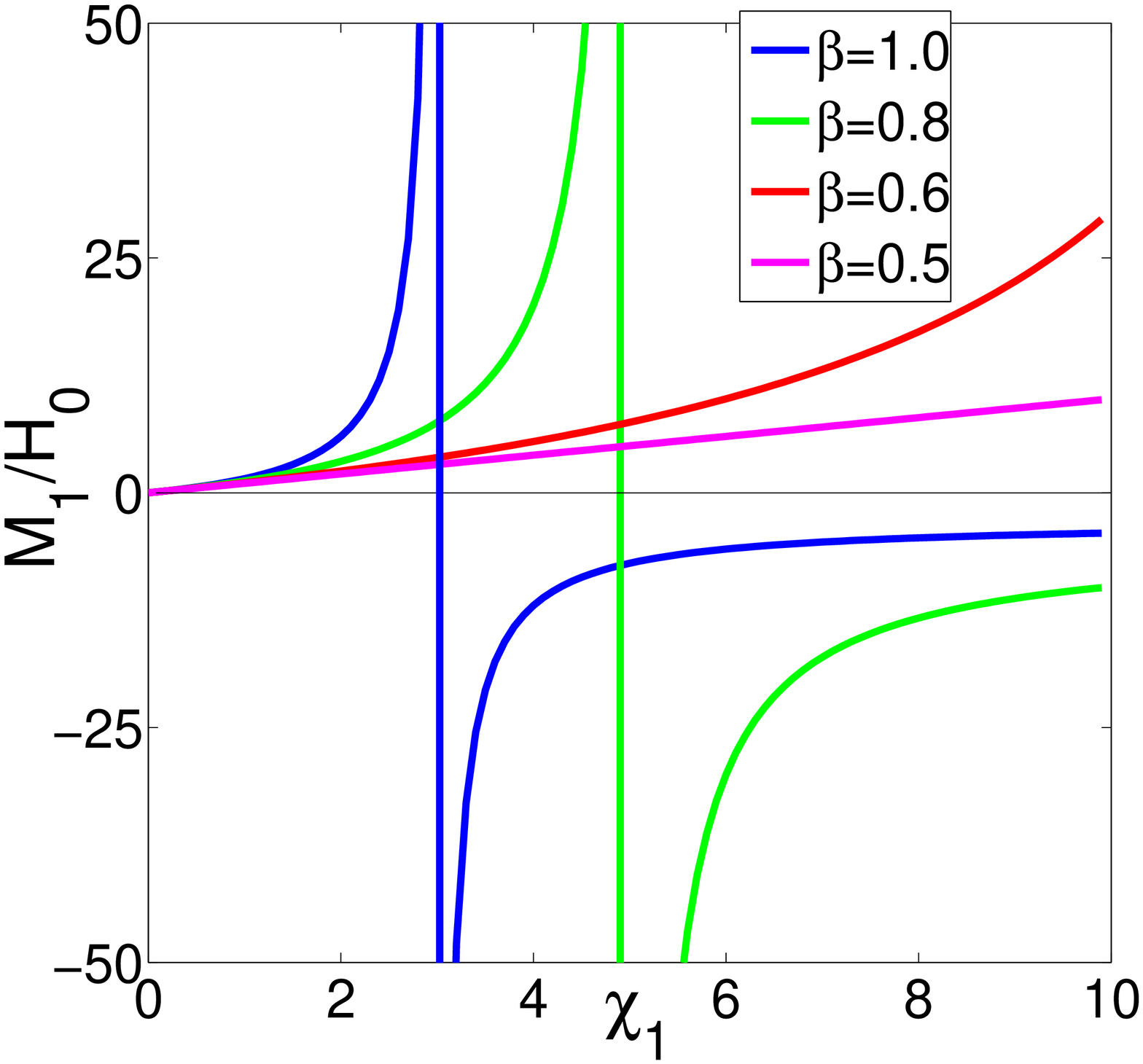}
\end{center}
\caption{(Color in online)  The magnetization $\vec M_1$ of layer {\it 1} depends on the parameter $\beta$. Other parameters:
$\chi_1 = \chi_2$, $\alpha_1 = \alpha_2=1/3$. 
 \label{fig-magnetization-pole}
}
\end{figure}

The meaning of the coefficient $\beta$ is obvious from the relation between the  magnetization
$M$ and the external field $H_0$ in magnetic gels: the magnetization should have the same direction as the applied field. As 
shown in Figure~\ref{fig-magnetization-pole}, where the magnetization $M_1$ of layer {\it 1} 
is plotted as a function of its susceptibility $\chi_1$ for the different values of $\beta$, at some $\beta$ a 
pole develops in Eq.(\ref{eq-9}). The pole causes a
nonphysical flipping over of the magnetization vector $\vec M_1$. Decreasing the value of $\beta$ 
guarantees the ``correct'' behavior of $\vec M_1$. All the setup configurations used in the main text 
are free from such pole effect for the values of $\beta \leq 1$. 
Physically, the factor $\beta$ contains the widening of the field lines away from the interfacial boundary of layer
 {\it 2}, see Figure~\ref{fig-2}.

\clearpage
\newpage

\bibliography{allahyarov.bib}
\bibliographystyle{unsrt}

\end{document}